\documentclass[11pt]{article}
\usepackage{adjustbox}
\usepackage{amsmath} 
\usepackage{amssymb} 
\usepackage{amsthm} 
\usepackage{amsfonts}

\usepackage{algorithm}  
\usepackage{algorithmicx}  
\usepackage{algpseudocode}  
\usepackage{setspace}

\usepackage{anyfontsize}
\usepackage{authblk}
\usepackage{apacite}
\usepackage[T1]{fontenc}
\usepackage{titlesec}
\usepackage{titletoc}
\usepackage[title]{appendix}
\usepackage{bm}
\usepackage{booktabs} 
\usepackage{caption}
\usepackage{cite}
\usepackage{color}
\usepackage[table,xcdraw]{xcolor}
\usepackage{dcolumn}
\usepackage{diagbox}
\usepackage{enumitem}
\usepackage{enumerate}
\usepackage{epstopdf}
\usepackage{float} 
\usepackage[perpage,symbol*]{footmisc}
\usepackage[T1]{fontenc}
\usepackage{graphics}
\usepackage{graphicx} 
\usepackage{geometry}
\usepackage{footmisc}
\usepackage{lipsum}
\usepackage{makecell}
\usepackage{mathrsfs}
\usepackage{multirow} 
\usepackage{natbib}
\usepackage{rotating}
\usepackage{setspace}
\usepackage{subfigure}
\usepackage{tabu}
\usepackage{tabularx}
\usepackage{threeparttable}
\usepackage{verbatim}
\usepackage{url}
\usepackage[colorlinks,urlcolor=blue,linkcolor=black,citecolor=blue]{hyperref}

\begin{document}

\title{\Large Optimal Bailout for Systemic Risk: \\ A PGO Approach Based on Neural Network\footnote{The work was supported by the National Natural Science Foundation of China (Nos. 72461160315, 72271250, 72371253, 72201074)
}}

\author[a]{Shuhua Xiao}
\author[b]{Jiali Ma}
\author[c]{Li Xia}
\author[c]{Shushang Zhu \thanks{\small{Corresponding author. School of Business, Sun Yat-Sen University, Guangzhou 510275, P.R. China. \\
E-mail addresses:
\href{mailto:shxiao3-c@my.cityu.edu.hk}{shxiao3-c@my.cityu.edu.hk}(S. H. Xiao), 
\href{mailto:majli@mail2.sysu.edu.cn}{majli@mail2.sysu.edu.cn}(J. L. Ma),
\href{mailto:xiali5@sysu.edu.cn}{xiali5@sysu.edu.cn}(L. Xia), 
\href{mailto:zhuss@mail.sysu.edu.cn}{zhuss@mail.sysu.edu.cn}(S. S. Zhu, corresponding author).
}}}

\affil[a]{\small{College of Business, City University of Hong Kong, Hong Kong, China}}
\affil[b]{\small{College of Big Data Statistics, Guizhou University of Finance and Economics, Guizhou 550025, P.R.China.}}
\affil[c]{\small{School of Business, Sun Yat-Sen University, Guangzhou 510275, P.R. China.}}

\date{}
\maketitle

\begin{abstract}
    In the financial system, bailout strategies play a pivotal role in mitigating substantial losses resulting from systemic risk. However, the lack of a closed-form objective function to the optimal bailout problem poses significant challenges in its resolution. This paper conceptualizes the optimal bailout (capital injection) problem as a black-box optimization task, where the black box is modeled as a fixed-point system consistent with the E-N framework for measuring systemic risk in the financial system. To address this challenge, we propose a novel framework, ``Prediction-Gradient-Optimization'' (PGO). 
    Within PGO, the \textit{Prediction} employs a neural network to approximate and forecast the objective function implied by the black box, which can be completed offline; For the online usage, the \textit{Gradient} step derives gradient information from this approximation, and the \textit{Optimization} step uses a gradient projection algorithm to solve the problem effectively.
    Extensive numerical experiments highlight the effectiveness of the proposed approach in managing systemic risk.
\end{abstract} 


\newpage
\section{Introduction}\label{sc:intro}
The intricate interconnections within the banking system can cause systemic risk \citep{diamond2005liquidity}. The 2008 Global Financial Crisis (GFC) exemplifies how a localized shock can spread contagiously through the financial network, ultimately triggering systemic disruptions across both the financial system and the broader economy. Preserving financial stability is therefore essential to minimize adverse spillovers to the real economy \citep{Bernanke1983effect}. Timely and well-designed rescue interventions are critical in this regard \citep{acharya2008cash}\footnote{In this paper, the terms ``bailout'' and ``rescue'' are used interchangeably.}. However, regardless of the specific mechanism, bailouts are invariably costly. Since the GFC, EU member states have committed nearly half a trillion euros in taxpayer-funded bailouts \citep{klimek2015bail}. These fiscal burdens naturally prompt the search for optimal bailout strategies under budgetary constraints.

Several studies have investigated optimal bailout implementation in various financial settings. \citet{aghion1999optimal} analyze the timing and instruments of efficient bailouts. \citet{leitner2005financial}, building on the framework of \citet{allen2000financial}, show that interbank linkages can be leveraged to incentivize private sector bailouts. In contrast, our focus is on identifying the optimal bailout strategy within a different financial network architecture. A seminal contribution in this area is the model of \citet{eisenberg2001systemic}, which introduces the so-called E–N framework—a network of interbank liabilities modeled as a fixed-point system—and demonstrates the existence of a clearing payment vector. This foundational model has since been extended in multiple directions \citep{rogers2013failure, amini2016fully, schuldenzucker2016clearing, chen2016optimization, csoka2018decentralized, kusnetsov2019interbank}. \citet{feinstein2017financial} incorporate multiple external assets and establish the joint existence of a clearing payment vector and a price vector. Building on this, \citet{ma2021joint} formally analyze the interaction between interbank liability networks and portfolio overlaps.

Admittedly, some researchers have examined bailout strategies within the context of E-N-type models. 
While the original E–N model can be cast as a linear program (LP), facilitating tractable solutions,
extensions that incorporate more realistic features complicate the characterization of optimal bailout strategies due to the loss of key structural properties. 
\cite{jackson2024credit} demonstrate that determining the minimum capital injections required to ensure solvency in a non-best equilibrium is strongly NP-hard. A central source of this complexity lies in the interdependence of solvency: the capital needed to make one bank solvent depends on which other banks are already solvent. Hence, the sequence of bailouts matters, as each round of intervention alters the balance sheets across the network.
To address these complexities, \citet{demange2018contagion} propose using aggregate debt repayments as a benchmark to guide intervention strategies. They define an institution's threat index as the system-wide spillover effect of a marginal increase in its cash injection on total debt repayments. Thus, this index serves as a tool for identifying optimal intervention targets.
\citet{dong2021some} propose a binary E–N model with budget constraints and show that the associated optimization problem is also NP-hard. They develop a sequential coefficient strengthening algorithm with proven global convergence guarantees. In another line of work, \citet{ma2021joint} frame the bailout problem as a mathematical optimization task and demonstrate that, although the objective lacks a closed-form expression, its gradient can be computed via implicit function theory. They subsequently employ a successive linear programming approach to solve it. 
In sum, under the E–N framework, the optimal bailout problem becomes significantly more complex as the model incorporates more realistic features. Extended E–N models often lead to NP-hard problems or lack closed-form objective functions, posing substantial challenges for optimization.

In this paper, we formulate the optimal bailout problem as a special case of a black-box optimization (BBO) problem. A black-box system is one in which outputs can be observed for given inputs, but the internal mechanisms that generate those outputs are analytically intractable \citep{audet2017derivative}. There are two primary perspectives in the analysis of black-box models: one aims to interpret or explain the underlying mechanism of the black box \citep{guidotti2018survey}, while the other seeks to approximate the black-box function through surrogate models \citep{vu2017surrogate}. In principle, once a sufficiently accurate approximation of the objective function is obtained, the original problem can be reframed as a conventional optimization task, which has been extensively studied in the literature.
Fortunately, the universal approximation theorem \citep{cybenko1989approximation, hornik1989multilayer} establishes that a neural network with at least one hidden layer and a sufficiently large number of neurons can approximate any continuous function to an arbitrary degree of accuracy. This theoretical result justifies the use of neural networks as surrogates for black-box functions, providing a foundation for the development of data-driven bailout optimization algorithms.

\begin{figure}[H]
  \centering
  \includegraphics[width=0.75\textwidth]{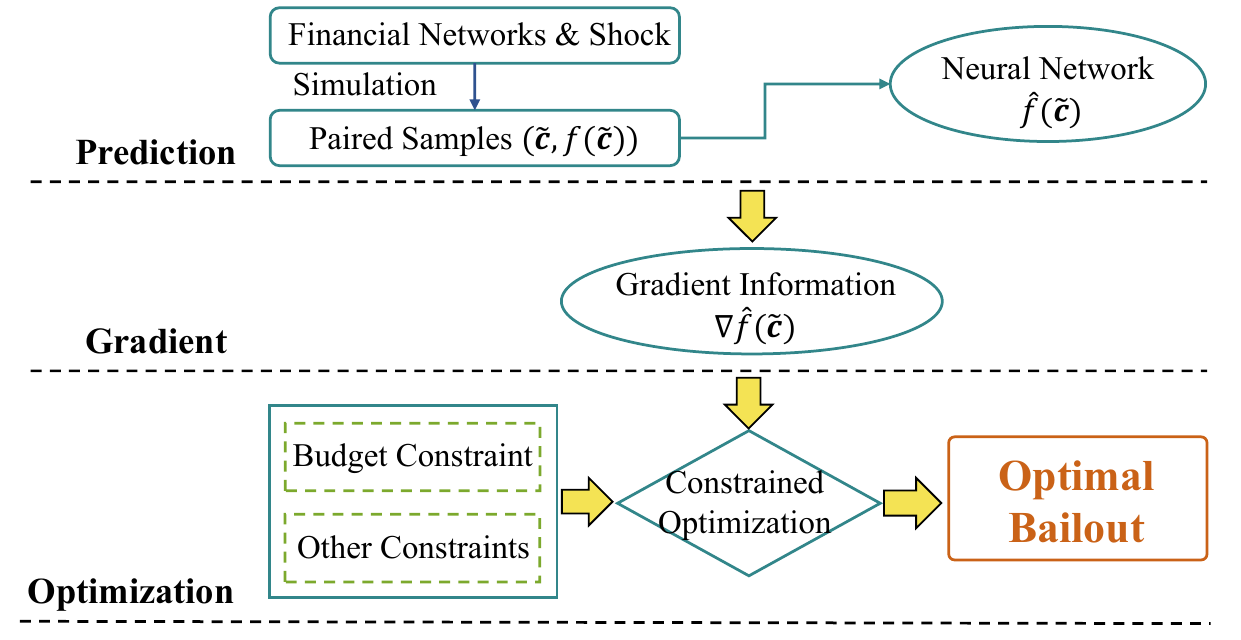}
  \caption{\small{\textbf{Roadmap for solving the optimal bailout problem.} 
  To train the neural network, we first generate simulated data consisting of pairs $\left( \tilde{\bm c}, f\left(\tilde{\bm c}\right) \right)$, where $\tilde{\bm c}$ denotes a candidate bailout strategy. Specifically, we randomly allocate bailout funds across banks, resulting in a corresponding payment vector and price vector for the financial system. These vectors are then used to evaluate the effectiveness of the bailout, producing labeled samples that serve as training data for the neural network.
  Once trained, the neural network provides gradient information, $\nabla f\left(\tilde{\bm c}\right)$. Equipped with an explicit functional form, gradient estimates, and constraint conditions, we apply standard constrained optimization methods to solve the problem. In doing so, we effectively transform a problem with an intractable objective function into a more manageable optimization task.
  }}
\label{fig:logic}
\end{figure}

Specifically, we propose a ``Prediction-Gradient Optimization'' (PGO) framework to solve the bailout problem. Figure~\ref{fig:logic} outlines the core idea. A neural network is employed to approximate the objective function that captures the effectiveness of rescue. We then apply backpropagation to compute the gradient of the objective with respect to the decision variables, i.e., the bailout vector. By combining the fitted objective and its gradient, we conduct a structured search for the optimal solution. It is important to emphasize that the neural network not only approximates the objective function but also provides essential gradient information, enabling the use of standard constrained optimization techniques.

Although random search may appear to be the simplest and most direct approach, the PGO framework offers several notable advantages. First, in high-dimensional banking systems, PGO consistently delivers effective solutions, whereas random sampling often requires extensive time to yield feasible outcomes. Second, real-world rescues typically face stringent budget constraints, making it difficult to generate admissible samples through a random search. In contrast, PGO naturally accommodates such constraints within the optimization process. Third, the flexibility of PGO allows it to address rescue problems under varying funding levels, offering a valuable tool for policymakers to evaluate trade-offs between costs and effectiveness.

The remainder of the paper is organized as follows. Section 2 introduces the research problem that motivates our framework. Section 3 outlines the principles and implementation details of the proposed PGO approach, including its application to practical bailout problems. In Section 4, we present numerical simulation results and compare them with those obtained from alternative methods. Section 5 concludes.

\section{Optimal Bailout: What Is It?}\label{sc:what}
In this section, we clarify the concept of optimal bailout by providing a detailed exposition of two financial system models: the original E–N model introduced by \citet{eisenberg2001systemic}, and the extended E–N model developed by \citet{feinstein2017financial}. We further illustrate how shocks propagate, how bailouts are implemented, and how bailout effectiveness is measured within each framework.

\subsection{Defining Financial Systems} \label{sub:ENLP}
The following analysis utilizes several standard concepts and definitions, which are consolidated to minimize confusion and facilitate referencing. Throughout the paper, $\mathbb{R}^{n}$ represents the $n$-dimensional Euclidean vector space, and $|\cdot|$ denotes the Euclidean norm. The bold lowercase letters represent vectors, and the uppercase letters represent matrices. Let $\bm{1}$ denote an $n$-dimensional vector whose components are all equal to 1. Similarly, let $\bm{0}$ denote an $n$-dimensional vector whose components are all equal to 0. For any two vectors $\bm{a}, \bm{b} \in \mathbb{R}^n$, the following operations are defined:
\begin{eqnarray*}
\bm a^{+}          &\triangleq &(\max\{a_{1},0\},\max\{a_{2},0\},\cdots,\max\{a_{m},0\})^{T}, \\
\bm a \wedge \bm b &\triangleq &(\min\{a_1,b_1\},\min\{a_2,b_2\}, \cdots, \min\{a_m,b_m\})^T, \\
\bm{f}(\bm a)      &\triangleq &(f_1(a_1),f_2(a_2),\cdots, f_m(a_m))^T.
\end{eqnarray*}

As noted in the Introduction, numerous models have been developed to characterize financial networks. In this paper, we adopt two of the most prominent frameworks. The first is the E–N model, a seminal contribution of \citet{eisenberg2001systemic}, which captures interbank liabilities and has attracted considerable attention in the literature. The second is the model proposed by \citet{feinstein2017financial}, which extends the E–N framework by incorporating multiple external assets. For brevity, we refer to this as the \textit{extended E–N model}.

\subsubsection{E-N Model}\label{sec:ENmodel}
Consider a financial system comprising $n$ banks indexed by $\mathcal{N}=\left \{ 1,\cdots,n \right \}$, where each bank can lend to or borrow from other banks, forming an interbank network. Here, the bank can be interpreted as any kind of financial institution. The liability relationships among banks are captured by the matrix $L=\left(l_{ij}\right)\in \mathbb{R}^{n\times n}$, where $l_{ij}$ denotes the nominal liability of Bank $i$ to Bank $j$. By construction, all liabilities are nonnegative and no bank lends to itself, i.e., $l_{ij} \geq 0$ and $l_{ii}= 0$, for any $i,j\in\mathcal{N}$. 
Let $\bar{\bm l}=(\bar{\bm l_i})_{n\times 1}$ denote the vector of total nominal obligations, where each component is defined as: $\displaystyle{\bar{l_{i}}=\sum\limits_{j=1}^{n}l_{ij}+b_{i}}$, with $b_{i}\in\mathbb{R}$ representing Bank $i$'s external liabilities, i.e., obligations to entities outside $\mathcal{N}$. Hence, $l_{ij}$ reflects interbank liabilities, while $b_{i}$ captures external obligations.
Let $\Pi=(\pi_{ij})\in \mathbb{R}^{n\times n}$ denote the relative liability matrix where $\pi_{ij}$ is defined by $\displaystyle{\pi_{ij} =\frac{l_{ij}}{\bar{l}_{i}} } $ if $\bar{l}_{i}>0$ and $\pi_{ij} =0$ otherwise.
Let $c_{i}\in\mathbb{R}$ denote the amount of cash held by Bank $i$, $e_i \geq 0$ the equity of Bank $i$. The balance sheet identity for Bank $i$ is then given by:
\begin{eqnarray}\label{eq:relationship1}
    w^{0}_{i}=\sum\limits_{j=1}^{n}\pi_{ji}\bar{l}_{j}+c_{i}=\bar{l}_{i}+e^{0}_{i},
\end{eqnarray}
where $w^{0}_{i}$ and $e^{0}_{i}$ denote the initial total asset and equity of Bank $i$, respectively. The initial state of the financial system can be uniquely parameterized as a triple $(\Pi, \bar{\bm{l}}, \bm{c})$.

Next, let us introduce the concept of the clearing vector, which specifies payments that adhere to certain rules among banks within a financial system. Specifically, there are three rules proposed by \cite{eisenberg2001systemic}: limited liabilities, the priority of debt claims over equity, and proportionality. In this regard, a clearing vector for the financial system $(\Pi, \bar{\bm{l}}, \bm{c})$ is a fixed point, $\bm{l}^{*} \in \left[0,\bar{\bm{l}}\right]$, of the map $\displaystyle{\Phi(\cdot ; \Pi, \bar{\bm{l}}, \bm c):[\bm{0}, \bar{\bm{l}}] \rightarrow[\bm{0}, \bar{\bm{l}}]}$ defined by:
\begin{equation}\label{eq:E-N vector}
    \Phi(\bm l ; \Pi, \bar{\bm l}, \bm c) \equiv\left(\Pi^{T} \bm l+ \bm c \right) \wedge \bar{\bm l}.
\end{equation}
By formulating the clearing mechanism of a financial system as a fixed-point problem, \citet{eisenberg2001systemic} establish the existence of a clearing payment vector and demonstrate that it can be computed via an iterative algorithm. Moreover, for any increasing function $f: \mathbb{R}^n \rightarrow \mathbb{R}$, they show that the clearing vector solves the following program:
$$\begin{array}{ll}
    \max \limits_{\bm l} & f(\bm l) \\
    \text { s.t. }       & \bm c+\Pi^{T}\bm l \geq \bm l  \enspace. \\
                         & \boldsymbol{0} \leq \bm l \leq \overline{\bm l}
\end{array}$$
Moving one more step, we can transform this problem into a linear programming problem by setting the objective function as
$f(\bm l)= \bm 1^{T}(\bm l-\bar{\bm l})$ or $f(\bm l) = \bm 1^{T}\bm l$ because the value of $\bm 1^{T} \bar{\bm l}$ is known.

\subsubsection{Extended E-N Model}
\begin{figure}[htbp]
    \centering
    \includegraphics[width=0.70\textwidth]{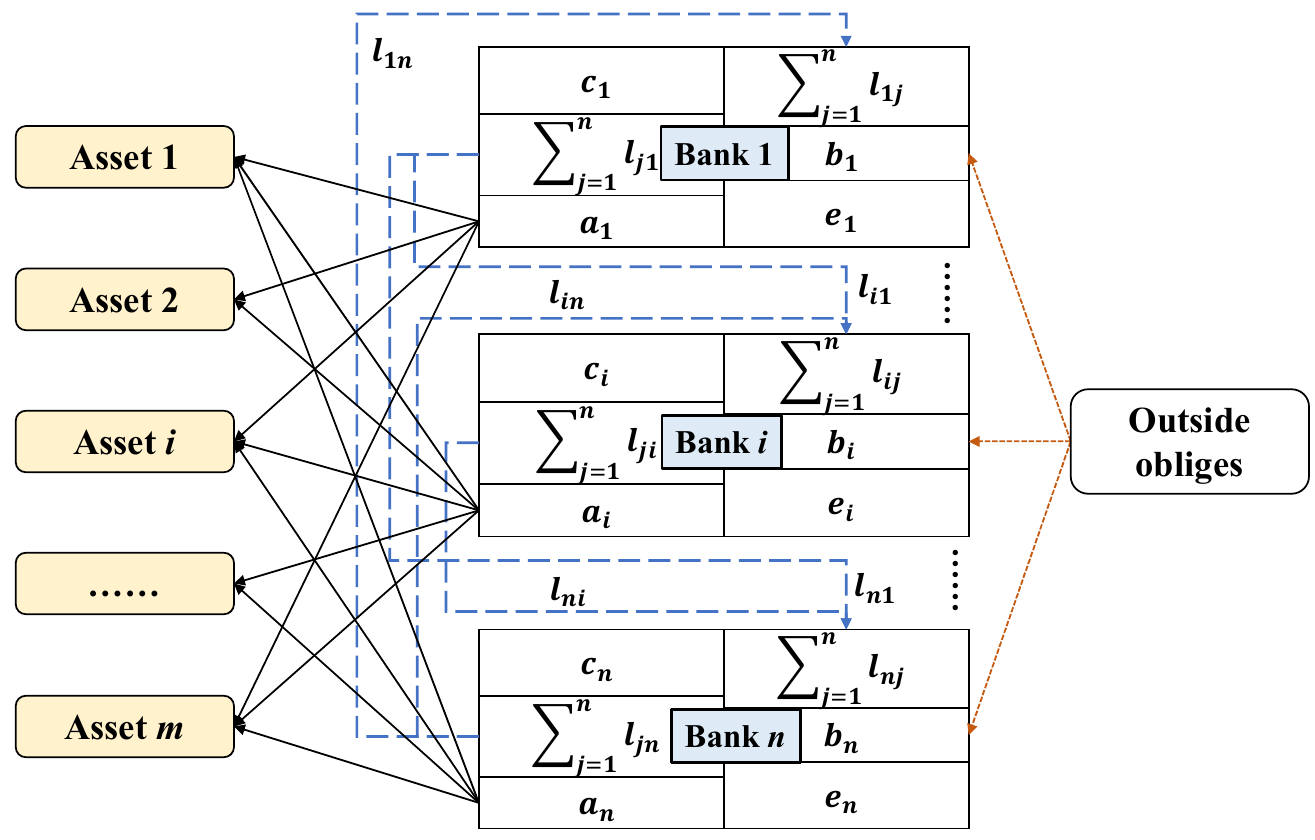}
    \caption{\small{\textbf{Banking System (Extended E-N Model).} 
    This figure shows the balance sheets of the banking system according to the extended E-N model. The equilibrium clearing system of the extended model can be defined in a similar way to that of the E-N model. We left the details in the sequel to avoid duplication.
    }}  
    \label{fig:banking-system}
\end{figure}

The primary distinction between the original E–N model and the extended E–N model lies in the incorporation of multiple external assets. Consider a financial system consisting of $n$ banks and a financial market comprising $m$ assets. Banks may lend to and borrow from each other, incur external liabilities, hold cash, and invest in the $m$ assets.
Let $A = (a_{ij}) \in \mathbb{R}^{n \times m}$ denote the portfolio holdings matrix, where $a_{ij}$ represents the quantity of asset $j$ held by Bank $i$. Without loss of generality, we assume that the initial price of each asset is normalized to 1, and that all banks are solvent at the initial time (time 0). In this setting, the asset-liability relationship for Bank $i$, as represented in Eq.~\eqref{eq:relationship1}, can be described as follows:

\begin{eqnarray*}
    w^{0}_{i}=\sum\limits_{j=1}^{n}\pi_{ji}\bar{l}_{j}+\sum\limits_{j=1}^{m}a_{ij}+c_{i}=\bar{l}_{i}+e^{0}_{i}.
\end{eqnarray*}
Similarly, the initial state of the financial system can be uniquely parameterized as a quadruple $(\Pi, \bar{\bm{l}},A,\bm{c})$. Figure~\ref{fig:banking-system} provides a schematic representation of the extended model.

\subsection{Responding to Shocks}
Suppose that an initial shock $\bm{s}=(s_i)\in\mathbb{R}_+^n$ occurs in the system, where $\displaystyle{s_{i}\geq 0}$ denotes the shock on Bank $i$. Such a shock may render certain banks insolvent, preventing them from fully repaying their interbank obligations. This, in turn, may trigger cascading defaults through the network, potentially resulting in asset losses or the insolvency of additional institutions and ultimately leading to systemic liquidation. Insolvent banks use their remaining assets to partially repay their creditors, while solvent banks fulfill their obligations in full. According to the principle of ``limited liability'', the asset $w_i$ used by Bank $i$ to pay its debts should be non-negative.

Fundamentally, the occurrence of a shock $\bm{s}$ reduces the cash holdings of the banks such that $\bm{c}:= \bm{c} - \bm{s}$. As a result, the state of the system --- represented by $(\Pi, \bar{\bm{l}}, \bm{c})$ in the E–N model or $(\Pi, \bar{\bm{l}}, A, \bm{c})$ in the extended E–N model --- is altered. For systems governed by the E–N model, the clearing vector can be recomputed using Eq.~\eqref{eq:E-N vector}. In contrast, for systems described by the extended E–N model, the inclusion of external assets substantially increases the complexity of the clearing process. In the following section, we focus on characterizing the joint response of financial institutions and asset markets to such shocks.

Faced with shocks, some banks are forced to sell their assets at fire-sale prices, and some of them may default on their obligations. Let $p_j$ denote the actual price of asset $j$ after fire sale. The liquidity strategy of Bank $i$ as a function of $(\bm l, \bm p)$, i.e., $\displaystyle{\bm{\eta}_{i}(\bm{l},\bm{p})=(\eta_{i1}(\bm{l},\bm{p}),\eta_{i2}(\bm{l},\bm{p}),\ldots\eta_{im}(\bm{l},\bm{p}))^{T}}$, where $\eta_{ij}(\bm{l},\bm{p})$ represents the amount of asset $j$ sold by Bank $i$ for given $\bm{l}$ and $\bm{p}$. Without loss of generality, according to \cite{feinstein2017financial} and \cite{ma2021joint}, we define $\eta_{ij}(\bm{l},\bm{p})$ as follow:
\begin{eqnarray*}
    \eta_{ij}(\bm{l},\bm{p})=\frac{a_{ij}}{\sum_{j=1}^{m}a_{ij}p_{j}}\left(\bar{l}_{i}-c_{i}-\sum_{k=1}^{n}\pi_{ki}l_{k}\right)^{+}.
\end{eqnarray*} 
It is evident that the liquidity strategy $\bm{\eta}_{i}(\bm{l},\bm{p})$ follows the proportional rule and is continuous and non-increasing in $(\bm{l},\bm{p})$.

For asset $j$, an inverse demand function can be applied $f_j:\mathbb{R}_{+}\rightarrow[p_j^{\min},1]\subseteq\mathbb{R}_{+}$ to quantify the price depreciation according to the fire sale, where $p_{j}^{\min}\geq 0$ represents the minimum price. As stated in \cite{ma2021joint}, it is reasonable to assume that for each $j\in\{1,2,\ldots,m\}$, the inverse demand function $f_{j}(x)$ satisfies $f_{j}(0)=1$ and that $f_{j}(x)$ is continuous and non-increasing in $x\in[0,+\infty)$. Aftershocks, based on payments ${\bm l}=(l_1,l_2,\cdots, l_n)^T$ and asset prices ${\bm p}=(p_1,p_2,\cdots, p_m)^T$, the total asset value of Bank $i$ and the updated price of asset $j$ become:
\begin{align*}
    w_{i}=\sum\limits_{j=1}^{n}\pi_{ji}l_{j}+\sum\limits_{j=1}^{m}a_{ij}p_{j}+c_{i} \quad \mbox{and} \quad
    p_{j}=f_{j}\left(\sum_{i=1}^{n}\left(a_{ij}\wedge\eta_{ij}(\bm{l},\bm{p})\right)\right),
\end{align*}
respectively. Meanwhile, recalling the rules of ``limited liability'' and ``absolute priority'', 
the liability payment vector $\bm l$ satisfies the following formula for each $i\in \mathcal{N}$:
\begin{eqnarray*}
    l_{i}=\left(\sum\limits_{j=1}^{n}\pi_{ji}l_{j}+\sum\limits_{j=1}^{m}a_{ij}p_{j}+c_{i}\right)^{+}\wedge \bar{l}_{i}.
\end{eqnarray*}

Now, the equilibrium clearing payment vector $\bm{l}$ and the price vector $\bm{p}$ can be depicted as a fixed point of the map $\Phi(\bm{l},\bm{p};\Pi,\bar{\bm{l}},A,\bm{c}):[\bm{0},\bar{\bm{l}}]\times[\bm{p}_{min},\bm{1}]\longrightarrow [\bm{0},\bar{\bm{l}}]\times[\bm{p}_{min},\bm{1}]$ parameterized by $(\Pi, \bar{\bm{l}},A,\bm{c})$ defined as:
\begin{eqnarray}\label{eq:F vector}
    \left(\begin{array}{c} \bm{l} \\ \bm{p}\\ \end{array} \right)
    =\Phi(\bm{l},\bm{p};\Pi,\bar{\bm{l}},A,\bm{c})\triangleq
    \left( \begin{array}{c} \left(\Pi^{T}\bm{l}+A\bm{p}+\bm{c}\right)^{+}\wedge\bar{\bm{l}}\\  
    \bm{f}\left(\sum_{i=1}^{n}\left(\bm{a}_{i}\wedge\bm{\eta}_{i}(\bm{l},\bm{p})\right)\right)\\
    \end{array} \right),
\end{eqnarray}
where $\bm a_i=(a_{i1},a_{i2},\cdots, a_{im})^T$.

Under certain mild conditions, such as the assumptions on $f_j$ and $\eta_{ij}$ described above, \cite{feinstein2017financial} proves the existence of the clearing vector. \cite{ma2021joint} propose a direct iterative algorithm based on the fixed-point system (\ref{eq:F vector}) and prove that the clearing vector, $(\bm{l}^*,\bm{p}^*)$, is a continuous function of the relative liability matrix $\Pi$, the nominal liability $\bm{\bar{l}}$, the asset holding matrix $A$ and the cash vector $\bm{c}$. 

Moreover, it is worth noting that the fixed-point equations used to define the clearing problem form a black-box system. This system operates by mapping inputs, such as injected funds or shocks, to outputs, such as payment vectors and price vectors, through iterative algorithms.
Except for the original E–N model discussed in Section~\ref{sec:ENmodel}, where the optimal bailout can be readily obtained, solving the optimal bailout problem in most other models poses considerable challenges. A significant portion of these models has been shown to be NP-hard (\cite{dong2021some,jackson2024credit}). In our extended E–N framework, the computational difficulty arises from the incorporation of cross holdings of financial assets --- a feature that, while adding complexity, reflects real-world structures and has been recognized as a critical channel for risk contagion during past financial crises. Thus, we aim to explore the optimal bailout strategy within this realistic setting.

\subsection{Measuring the Effectiveness of Bailouts}\label{sub:diffcult-optimal}
In reference to the bailout, it can be designated as $\bm{\tilde{c}}=(\tilde{c}_i)\in\mathbb{R}_+^n$, where $\displaystyle{\tilde{c}_{i}\geq 0}$ represents the fund injected into Bank $i$. Actually, its impact on the banking system is similar to that of $\bm s$. Hence, after shocks and bailouts, the state of the banking system is updated to $(\Pi, \bar{\bm{l}}, \bm c- \bm s+\tilde{\bm c})$ / $(\Pi, \bar{\bm{l}}, A, \bm c-\bm s+\tilde{\bm c})$, and the payment vector $\bm l^{*}$ and the price vector $\bm p^{*}$ are updated accordingly.

As indicated in Introduction, the key question we need to address is: What constitutes the optimal bailout? A related question is how to assess the efficacy of the bailout. In fact, the criteria for measuring effectiveness vary depending on the type of model. In the case of the E-N model, an appropriate measure of effectiveness is the maximum clearing vector after the bailout, denoted by $Pay_{all}$ and defined as follows:
\begin{equation}\label{eq:payall}
    Pay_{all} = \bm{1}^{T}\bm{l}^{*}(\bm{\tilde{c}}),
\end{equation}
where $\bm{l}^{*}(\bm{\tilde{c}})$ is the payment vector associated with the bailout $\bm{\tilde{c}}$ after the shock.

In the extended E–N model, following the formulation in \citet{ma2021joint}, the effectiveness of a bailout is evaluated using the metric $Save_{all}$, which is defined as the sum of two components: $Save_{in}$ and $Save_{out}$.
Here, $Save_{in}$ represents the reduction in total losses incurred by financial institutions due to the bailout, while $Save_{out}$ captures the reduction in losses borne by external obligees. These components are formally defined as follows:
\begin{align}
    Save_{in~}&= \tilde{\bm c} + \Pi^{T}\left[\bm l^{*}\left(\tilde{\bm c}\right) - {\bm l^{*}}(\bm s)\right] + A\left[\bm p^{*}(\tilde{\bm c})-\bm p^{*}(\bm s)\right], \label{eq:savein} \\ 
    Save_{out}&=\left( \bm{1}^{T}-\bm{1}^{T}\Pi \right) \left[\bm l^{*}\left(\tilde{\bm c}\right) - {\bm l^{*}}(\bm s)\right] ,  \label{eq:saveout} \\ 
    Save_{all}&=\tilde{\bm c} + \Pi^{T}\left[\bm l^{*}\left(\tilde{\bm c}\right) - {\bm l^{*}}(\bm s)\right] 
               +\left( \bm{1}^{T}-\bm{1}^{T} \Pi \right) \left[\bm l^{*}\left(\tilde{\bm c}\right) - {\bm l^{*}}(\bm s)\right] 
               +A\left[\bm p^{*}(\tilde{\bm c})-\bm p^{*}(\bm s)\right], \label{eq:saveall}
\end{align}
where $\bm{l}^{*}(\bm{\tilde{c}})$  and $\bm{p}^{*}(\bm{\tilde{c}})$ is the payment vector and the price vector after the shock and the bailout, respectively; $\bm{l}^{*}(\bm{s})$ and $\bm{p}^{*}(\bm{s})$ are the payment vector and the price vector after the shock, respectively. 

The objective of the bailout is to maximize one or more of the above criteria. 
It should be noted that the values of $\bm{l}^{*}(\cdot)$ and $\bm{p}^{*}(\cdot)$ must be computed iteratively using a fixed-point system. Therefore, although the values of $Pay_{all}$, $Save_{in}$, $Save_{out}$ and $Save_{all}$ can be derived using Eq.(\ref{eq:payall}), Eq.(\ref{eq:savein}), Eq.(\ref{eq:saveout}) and Eq(\ref{eq:saveall}), respectively, they do not possess closed-form expressions.

In addition, \cite{ma2021joint} also propose a more intuitive indicator $Ratio$ defined as follows:
\begin{equation}\label{eq:ratio}
    Ratio = \frac{Save_{all}}{\tau},
\end{equation}
where $\tau$ is the bailout budget. Notice that $Ratio$ can directly measure the efficiency of the bailout in the relative sense.

It is evident that the evaluation criteria mentioned earlier are non-concave functions in relation to the injected fund $\bm{\tilde{c}}$, and it is easy to provide examples to illustrate this fact. Moreover, there is no explicit or analytic representation between the criteria and the injected fund $\tilde{\bm c}$. The functions $\bm l^{*}\left(\tilde{\bm c}\right)$ and $\bm p^{*}(\tilde{\bm c})$ can only be evaluated using numerical algorithms, such as the iterative algorithm proposed by \cite{ma2021joint}. Therefore, conventional optimization methods cannot be applied directly to identify the optimal bailout.

\section{Optimal Bailout: How to Find It?}\label{sc:how}
In this section, we introduce the principles behind our PGO method and outline the cases that we aim to address, enabling us to apply the PGO method to solve these cases in the next section.

\subsection{PGO: Prediction, Gradient and Optimization}\label{se:pgo}
\begin{figure}[H]
    \centering
    \includegraphics[width=0.7\textwidth]{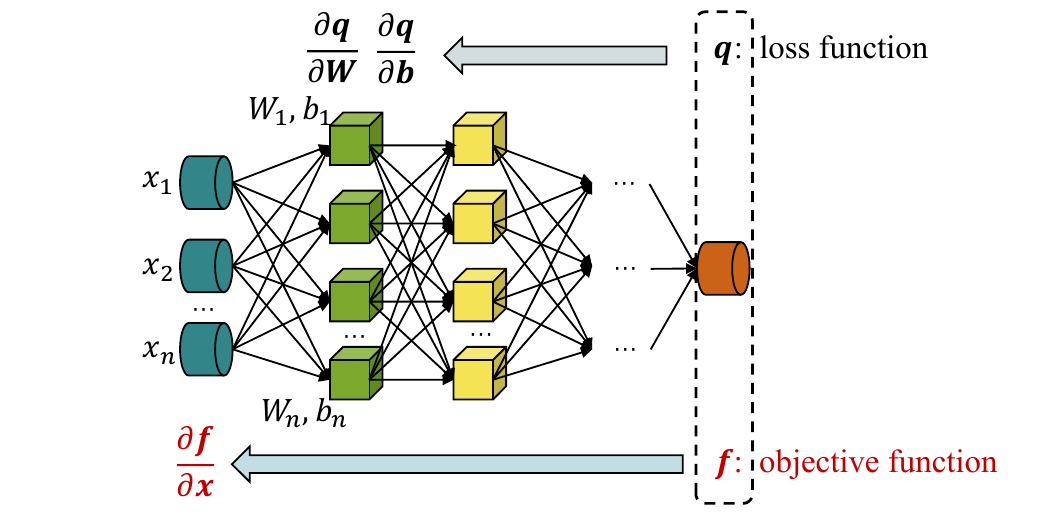}
    \caption{\small{\textbf{Distinguishing the gradient information used in this paper from commonly used gradient information.} This figure shows the gradient information that we use. In general, the gradient information of the loss function $\bm q$ with respect to network weights $\bm W$ and biases $\bm b$ is used to optimize the loss function. However, in our PGO method, we further compute the gradient of the objective function $\bm f$ approximated with a neural network with respect to $\bm x$ .}}
    \label{fig:NN-different}
\end{figure}

In this subsection, we briefly outline the core principle of the PGO method. In our framework, a neural network is employed as a functional approximation tool, enabling the extraction of gradients necessary to optimize the bailout strategy. Specifically, we use the backpropagation algorithm to compute the gradient of the loss function with respect to the parameters of the neural network.
Our approach departs from existing literature in a key way: rather than focusing on parameter optimization, we leverage the trained neural network to compute the gradient of the fitted objective function with respect to its inputs, as illustrated in Figure~\ref{fig:NN-different}.
The detailed procedure for computing the gradient of the objective function—modeled via a neural network—is provided in Appendix~\ref{ap:gradient}. This approach is particularly well-suited to black-box optimization problems. Within such settings, a sufficiently large number of input–output samples can be generated from the black-box system to train a neural network that approximates the objective function. The resulting gradient evaluator, derived from the trained network, can then be incorporated into various gradient-based optimization algorithms.
Importantly, the systemic risk bailout problem can be rigorously framed as a black-box optimization task, where the fixed-point system defining the financial clearing mechanism serves as the black box.

In the following, we outline the neural network-based black-box optimization approach in a more general framework. Consider the following constrained optimization problem:
$$\begin{array}{lll}
    &\min          & f(\bm x) \\
(P) &\text {s.t.}  & g(\bm x) \leq 0  \enspace,\\
    &              & h(\bm x) =0
\end{array}$$
where the objective function $f(\boldsymbol{x})$ has no analytic expression, but it can be evaluated for each given $\bm x$ with a black-box system. Denote the approximation of $f(\boldsymbol{x})$ via the neural network as $\hat{f}(\boldsymbol{x})$, which is trained with a sufficiently large number of samples of $(\boldsymbol{x},f(\boldsymbol{x}))$ generated by the black box system. 

Upon obtaining the objective function predictor and the gradient calculator, we can solve $(P)$ with PGO, and the sketch is clarified as the following Algorithm \ref{alg:PGO}. 

\begin{algorithm}[H]
    \setstretch{1.25}
    \caption{Prediction-Gradient-Optimization (PGO)}
    \label{alg:PGO}
    \begin{algorithmic}[1] 
    \Require
        Training set of $(\boldsymbol{x},f(\boldsymbol{x}))$ generated by black-box system; Times of training $\mathcal{T}$; Inequality constraint $g(\boldsymbol{x})$; Equality constraint $h(\boldsymbol{x})$; Initial point $\boldsymbol{x}_0$
    \Ensure
        $\boldsymbol{x}^*$
    \State Initialize $\boldsymbol{W}$ and $\boldsymbol{b}$ randomly;
    \While{the times in $\mathcal{T}$}
        \State According to the optimization of the loss function based on $(\boldsymbol{x},f(\boldsymbol{x}))$, update $\boldsymbol{W}$ and $\boldsymbol{b}$;
    \EndWhile
    
    \Function{Prediction}{$\boldsymbol{W}$, $\boldsymbol{b}$, $\boldsymbol{x}$}
        \State According to the forward-propagation process, predict $\hat{f}(\boldsymbol{x})$;
        \State \Return{$\hat{f}(\boldsymbol{x})$};
    \EndFunction

    \Function{Gradient}{$\boldsymbol{W}$, $\boldsymbol{b}$, $\boldsymbol{x}$}
        \State According to Eq.(\ref{eq:deltauv}), compute $\nabla \hat{f}(\boldsymbol{x})$;
        \State \Return{$\nabla \hat{f}(\boldsymbol{x})$};
    \EndFunction

    \Function{Optimization}{$\hat{f}(\boldsymbol{x})$, $\nabla \hat{f}(\boldsymbol{x})$, $g(\boldsymbol{x})$, $h(\boldsymbol{x})$, $\boldsymbol{x}_0$}
        \State Starting from $\boldsymbol{x}_0$, use one constrained optimization algorithm to optimize $\boldsymbol{x}$;
        \State \Return{$\boldsymbol{x}^*$}.
    \EndFunction
\end{algorithmic}
\end{algorithm}

Note that Algorithm~\ref{alg:PGO} serves as a general-purpose PGO solver. In theory, this method is designed to identify locally optimal solutions to the optimization problem, rather than guaranteeing global optimality. The choice of optimization algorithm should therefore be guided by the specific structure of the problem at hand.
In the case of the systemic risk bailout problem, where all constraints are linear, we adopt the gradient projection algorithm (GPA) \citep{rosen1960gradient} as an efficient solution method. Although GPA has been extensively analyzed under constant step-size settings \citep{gafni1984two,calamai1987projected,ruszczynski2011nonlinear}, and several studies have explored adaptive step-size rules \citep{tseng1998incremental}, we opt for a simple yet flexible approach to selecting an acceptable step-size, tailored to the particular characteristics of our problem. The complete details of the implementation are provided in Appendix~\ref{ap:gradient projection}.

As a black-box optimizer that couples surrogate-model fitting with decision making, the algorithm must be assessed for rationality, convergence, and stability. While the proposed PGO framework is primarily data‐driven, classical results provide a high-level justification for these considerations. 
The rationality of the PGO method basically hinges on the ability of the neural network to accurately approximate the black-box objective function, such as $ Pay_{all} $ or $ Save_{all} $, which lacks a closed-form expression. According to the universal approximation theorem \citep{cybenko1989approximation, hornik1989multilayer}, in the context of our bailout problem, given a sufficiently large and representative training dataset of input-output pairs $ (\tilde{\bm c}, f(\tilde{\bm c})) $, the neural network can theoretically approximate the true objective function $ f(\tilde{\bm c}) $ and its derivatives with any precision. Notice that the surrogate-model fitting is separable to the decision making, thus training can be done offline with a large number of samples to ensure the accuracy of approximation. 

In the optimization phase, provided the surrogate model is sufficiently accurate, the gradient projection algorithm is known to converge to a stationary point for problems with differentiable objective functions and linear constraints, provided the step size is chosen appropriately \citep{rosen1960gradient,ruszczynski2011nonlinear}. In our implementation, we adopt a conservative step-size selection strategy (Appendix \ref{ap:gradient projection}) to ensure stable convergence, although this may not always yield the global optimum due to the non-convex nature of the objective function.

Stability in the PGO framework should mainly be interpreted as the robustness of the solution to perturbations in the input data and the parameters of the neural network. As described in Section \ref{sc:case1&2}, we generate training data by simulating various bailout scenarios, including valid and invalid samples, to mitigate overfitting and ensure robustness. In addition, stability is further enhanced by the continuity of the clearing vectors $ \bm{l}^*(\tilde{\bm c}) $ and $ \bm{p}^*(\tilde{\bm c}) $, as established by \citet{ma2021joint}, which ensures that small changes in the bailout vector $ \tilde{\bm c} $ result in bounded changes in the objective function.

\subsection{Two Practical Cases}
\begin{figure}[H]
    \centering
    \includegraphics[width=0.7\textwidth]{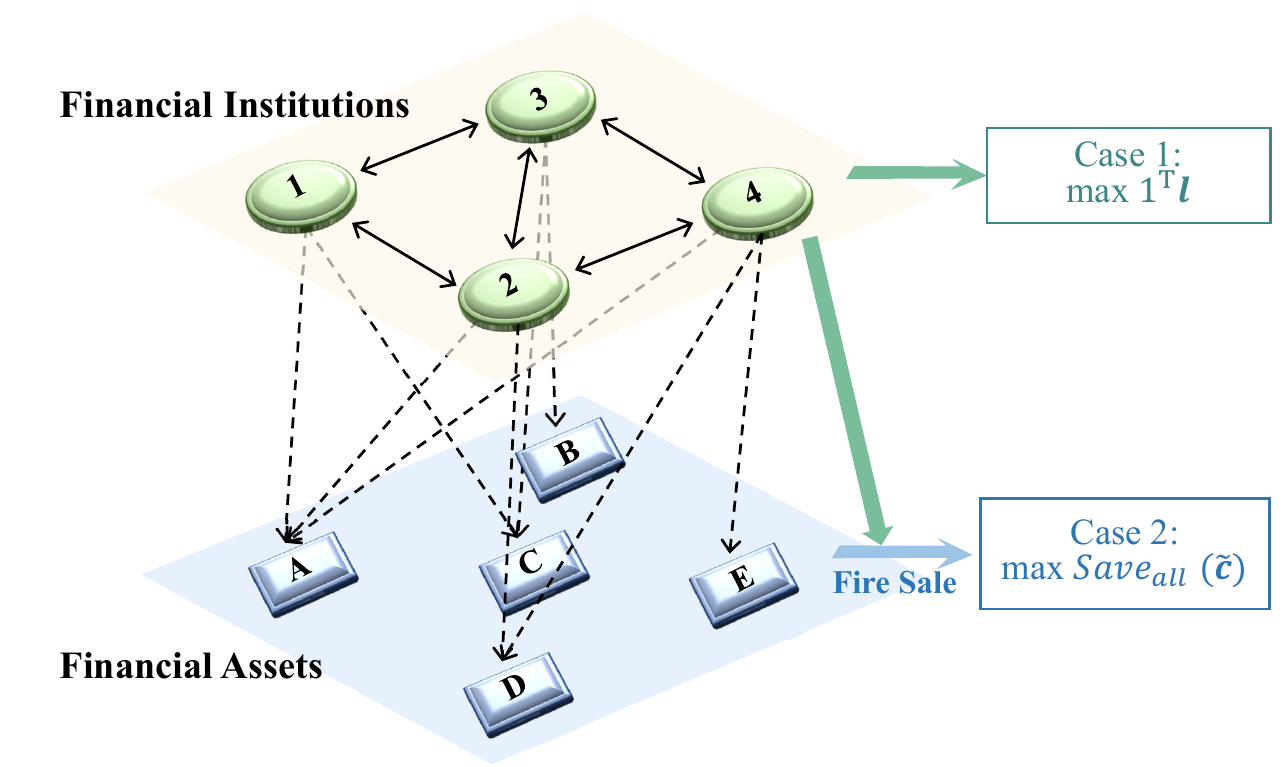}
    \caption{\small{\textbf{Relationship between cases.} The interdependence between cases can be categorized as follows: \textit{Case 1} examines solely the interbank relationships among financial institutions, whereas \textit{Case 2} scrutinizes both interbank relationships and portfolio overlapping relationships.}}
    \label{fig:problem}
\end{figure}

There are two practical bailout scenarios that correspond to the two models discussed earlier. In this subsection, we clarify the relationships and distinctions between these cases, highlighting their key properties, objective functions, decision variables, and potential solutions. Figure~\ref{fig:problem} offers a graphical representation of the connection between these two problems.

\textit{Case~1} is based on the original E–N model, which focuses on the interbank financial network. \textit{Case~2}, in contrast, builds on the extended E–N model, incorporating not only the interbank network but also allowing each bank to hold external assets and engage in fire sales in the event of a shock. These distinct configurations give rise to different characteristics in the two cases. In \textit{Case~1}, the solution for the payment and rescue vectors can be expressed as a linear program (LP) \citep{eisenberg2001systemic}, which is computationally tractable. However, in \textit{Case~2}, the inclusion of overlapping portfolios and fire sales significantly increases the complexity of solving the optimal bailout. The objective function in \textit{Case~2} cannot be explicitly expressed, rendering the problem much harder to solve.

How do we verify the effectiveness of the PGO approach? For \textit{Case~1}, we compare the PGO results with those obtained from the known optimal bailout solution, which is solved precisely by linear programming (LP). Specifically, we use the LINPROG function in MATLAB, which, by default, employs a dual-simplex algorithm to solve linear programming problems and obtain the optimal solution accurately. By comparing the gaps between the PGO solution and the optimal LP solution, we gain an intuitive understanding of the efficiency of the PGO approach.
For \textit{Case~2}, \citet{ma2021joint} proposed an effective heuristic algorithm based on successive linear programming, using the objective function $Pay_{all}$ rather than $Save_{all}$. We refer to this method as ``Heuristic'' for brevity. However, with the PGO method, we can optimize the objective function $Save_{all}$. To compare the efficiency of the two algorithms, we also derive the corresponding PGO results using $Pay_{all}$ as the objective function. 
Except for \textit{Case~1}, where both $\bm l$ and $\bm c$ can be treated as decision variables when solving via LP, in all other methods, $\bm c$ is decided first, and the clearing vector $\bm l$ is subsequently computed based on the outcome of the bailout. The reason for this is as follows.
On the one hand, in the extended E–N model, the banking system initially exists in an equilibrium state, and after receiving the bailout ($\bm{c}$), the system undergoes a new clearing process, ultimately reaching a new equilibrium. There is a dependency between $\bm l$ and $\bm c$, with $\bm l$ conditional on $\bm c$. On the other hand, PGO establishes the relationship between $\tilde{\bm c}$ and $Pay_{all}$ (or $Save_{all}$), and the details of the clearing process are not explicitly modeled.  Theoretically, it is possible to establish a relationship between $\tilde{\bm l}$ and $Pay_{all}$ (or $Save_{all}$). However, based on $\bm l$, there may be multiple solutions of $\tilde{\bm c}$, which means that the uniqueness is not guaranteed. What we can confirm is that the $\tilde{\bm l}$ derived from $\tilde{\bm c}$ is unique. Moreover, in real-world banking systems, the dynamics typically involves receiving a bailout first, followed by a new clearing process.
Table~\ref{tb:2cases} summarizes the circumstances described above.
In the ``Objective Function'' column of Table \ref{tb:2cases}, both the LINPROG and Heuristic methods adopt $Pay_{all}$ as the objective function due to limitations imposed by the model structure or the inaccessibility of implicit gradients. In contrast, the PGO method offers greater flexibility, allowing both $Pay_{all}$ and $Save_{all}$ to serve as objective functions. As shown in the ``Decision Variables'' column, only the LINPROG method can use both $\bm l$ and $\bm c$ as decision variables.

\begin{table}[H]\small
\centering
\caption{Brief summary of two cases.}
\renewcommand\arraystretch{1.25}
\setlength{\tabcolsep}{5.1mm}{
\begin{tabular}{cccccc}
\toprule
\textbf{Case} & \textbf{Model}   & \textbf{Method} & \textbf{Objective Function} & \textbf{Decision Variables} \\ \midrule
\multirow{2}{*}{\textit{Case 1}} &\multirow{2}{*}{E-N Model} & LINPROG & $Pay_{all}$  & $\bm l$, $\tilde{\bm c}$  \\
   && PGO               & $Pay_{all}$          & $\tilde{\bm c}$                 \\ \midrule
\multirow{3}{*}{\textit{Case 2}} & \multirow{3}{*}{\shortstack{Extended \\  \\ E-N Model}} & Heuristic  & $Pay_{all}$  & $\tilde{\bm c}$    \\
   && PGO               & $Pay_{all}$          & $\tilde{\bm c}$                 \\
   && PGO               & $Save_{all}$         & $\tilde{\bm c}$                 \\ \bottomrule
\end{tabular}}
\label{tb:2cases}
\end{table}

\section{Simulation Results}\label{sc:case1&2}
In this section, we employ the PGO method in the preceding two cases where the TensorFlow version used is 1.15.0. The computing device utilized to execute the programs is equipped with an Intel(R) Core(TM) i7-6700HQ CPU, 8.00GB RAM, and an NVIDIA GeForce GTX 960M.

\subsection{Case 1} 
In this case, banks encounter a shock represented by $\bm s$ and receive a bailout of $\tilde{\bm c}$. Based on the prior definition of $Pay_{all}$, the optimal bailout problem can generally be formulated as follows:
$$\begin{array}{ll}
\max \limits_{\tilde{\boldsymbol{c}}} & Pay_{all}(\tilde{\boldsymbol{c}}) \\
\text { s.t. } & \bm{1}^{T} \tilde{\boldsymbol{c}} \leq \tau  \enspace, \\
               & \tilde{\boldsymbol{c}} \geq \bm{0}
\end{array}$$
\noindent where $\tau$ is the total budget for the bailout. 

Following \cite{eisenberg2001systemic}, we formulate the problem in \textit{Case~1} as the following linear program:
$$\begin{array}{ll}
\max \limits_{\boldsymbol{l}, \tilde{\boldsymbol{c}}} & \bm{1}^{T} \boldsymbol{l} \\
\text { s.t. }  & \tilde{\boldsymbol{c}}+\boldsymbol{c}-\boldsymbol{s}+\Pi^{T} \boldsymbol{l} \geq \boldsymbol{l} \\
                & \bm{0} \leq \boldsymbol{l} \leq \overline{\boldsymbol{l}}   \qquad\qquad\enspace\enspace,     \\
                & \bm{1}^{T} \tilde{\boldsymbol{c}} \leq \tau                        \\
                & \tilde{\boldsymbol{c}} \geq \bm{0}
\end{array}$$
which can be solved with accuracy by LINPROG in MATLAB and is considered a benchmark for evaluating the efficiency of PGO. Notice that when $\tau=0 \left( \tilde{\bm c}=0\right)$, the problem is equivalent to computing the clearing vector in the E-N model. 
Here, the linear programming model that can be solved accurately and globally serves as a benchmark for comparison.

When using the PGO method to solve the optimal bailout as a black-box optimization, 
we first determine $\tilde{\bm c}$, then compute $\bm l$ through the fixed-point system,
and subsequently determine $Pay_{all}$. The effectiveness of PGO can be evaluated by comparing its results with those obtained from LINPROG: a close match between them would validate the efficiency of the PGO method.

We simulate the banking system as follows. We consider three different values for the number of financial institutions: 10, 100, and 1000. The total initial assets, $w_i^0$, of Bank $i$ are randomly drawn from a uniform distribution $U(0, 1)$. A randomly selected 10\% of the banks in the system are subjected to bankruptcy due to exogenous shocks. For the remaining parameters, we adopt the settings from \citet{ma2021joint}. Specifically, we use the Erd\H{o}s-Rényi (E-R) network \cite{erdHos1960evolution} as the liability network and set three key parameters of the banking system: inter-institutional leverage, degree of overlapping, and asset-liability ratio, to values of 0.7, 0.5, and 0.7, respectively. For further details on these parameters, see \citet{ma2021joint}.
\footnote{The neural network architecture is detailed in Appendix~\ref{ap:inf-nn}. Unless stated otherwise, each neural network is trained using 10,000 samples.}
The reason for using simulation data is twofold: first, interbank asset-liability relationships and joint holdings of financial assets are unavailable and must be simulated through assumptions; second, the number of banks with publicly available data is limited in practice, yet we need to comprehensively test our method across low, medium, and high-dimensional scenarios.

Table \ref{tb:LP-rescue} reports the performance of the LINPROG and PGO methods under \textit{Case~1}. Here, $n$ denotes the dimensionality of the banking system. The column labeled ``LINPROG'' presents the optimal benchmark values obtained by linear programming. The columns under ``PGO'' correspond to different depths of the neural network, with ``lay'' indicating the number of hidden layers --- larger values of ``lay'' represent deeper networks. For each configuration (e.g., $n=10$ and $lay=2$), two pieces of information are provided: the first is the value obtained by the PGO method, and the second is the percentage this value represents relative to the LINPROG benchmark. For example, a result of 3.979 and 95.98\% implies that when $n=10$ and $lay=2$, the PGO solution achieves 95.98\% of the optimal value obtained from LP.

The results highlight the effectiveness of the PGO approach and suggest that an appropriate number of hidden layers can enhance performance, while further increasing the number of layers does not necessarily yield better results.
As discussed in Section~\ref{se:pgo}, the choice of optimizer is flexible during the PGO optimization process. To explore this, we employ the SLSQP optimizer, a sequential least squares programming algorithm implemented in Scipy for Python, to solve the problem. The results are notably similar to those obtained with LINPROG. A detailed comparison of the results is provided in Appendix~\ref{ap:slsqp}.

\begin{table}[H]\small
\centering
\renewcommand\arraystretch{1.25}
\caption{Results of LINPROG and PGO in \textit{Case~1}.}
\setlength{\tabcolsep}{6mm}{
\begin{tabular}{cccccc}
\toprule
\multirow{2}{*}{$\bm{n}$} & \multirow{2}{*}{\textbf{LINPROG}}   & \multicolumn{4}{c}{\textbf{PGO}}      \\ 
\cline{3-6} 
&                         & \bm{$lay=2$} & \bm{$lay=3$} & \bm{$lay=4$}   & \bm{$lay=5$}   \\ 
\midrule
\multirow{2}{*}{$n=10$}   & \multirow{2}{*}{4.146}   & 3.979    & 4.059    & 4.136    & 4.046    \\
                          &                          & 95.98\%  & 97.89\%  & 99.76\%  & 97.58\%  \\ 
\midrule
\multirow{2}{*}{$n=100$}  & \multirow{2}{*}{28.859}  & 27.932   & 28.787   & 27.785   & 28.751   \\
                          &                          & 96.79\%  & 99.75\%  & 96.281\% & 99.63\%  \\ 
\midrule
\multirow{2}{*}{$n=1000$} & \multirow{2}{*}{312.853} & 311.030  & 311.852  & 311.553  & 311.936  \\
                          &                          & 99.42\%  & 99.68\%  & 99.58\%  & 99.71\%  \\ 
\bottomrule
\end{tabular}}
\label{tb:LP-rescue}
\end{table}

\subsection{Case 2} \label{sc:case2}
The case considered in this sub-section is based on the extended E-N model presented in \cite{feinstein2017financial}. As indicated in Section \ref{sub:diffcult-optimal}, the optimal bailout problem is significantly more challenging in this case because of the more intricate relationships that are characterized by the fixed-point system (\ref{eq:F vector}). This is particularly true for problems with the objective function $Save_{all}$.

Using $Pay_{all}$ rather than $Save_{all}$ as the objective function, \cite{ma2021joint} creatively propose a heuristic algorithm based on the following model:
$$\begin{array}{ll}
    \mathop{\max}\limits_{\tilde{\bm c}_{\mathcal{D}^{1}}} 
                   & Pay_{all}(\tilde{\bm c}_{\mathcal{D}^{1}}) = \bm{1}^{T} \bm l_{\mathcal{D}^{1}}^{*}\left(\tilde{\bm c}_{\mathcal{D}^{1}}\right) \\ 
    \text { s.t. } & \bm{1}^{T} \tilde{\bm c}_{\mathcal{D}^{1}} \leq \tau \enspace, \\ 
                   & \tilde{\bm c}_{\mathcal{D}^{1}} \geq \bm{0}
\end{array}$$
where $\mathcal{D}^{1}$ indicates the collection of banks that become insolvent but still have positive assets.

Notice that the bailout problem described above involves injecting cash only into banks in $\mathcal{D}^{1}$, which are classified as partially defaulted institutions. It does not consider fully defaulted institutions with non-positive assets. While this restriction may lead to suboptimal solutions, it facilitates the calculation of the gradient via the implicit function theorem, thus allowing the optimal bailout problem to be solved using gradient descent methods \citep{ma2021joint}.
Furthermore, while using $Pay_{all}$ as the objective function is effective, it is a reluctant compromise, constrained by the gradient calculation process. If $Save_{all}$ were selected as the objective function, the calculation of the gradient would become infeasible.

The PGO method offers greater flexibility and is not limited by the constraints discussed earlier. Recalling Eq.~\eqref{eq:saveall}, it is clear that $Save_{all}$ serves as an objective function that accounts for the reduction in total losses incurred by both internal and external obligees, thereby providing a more comprehensive assessment of the overall impact of the bailout on the financial system. In light of these considerations, we proceed to formulate the following problem:
$$\begin{array}{ll}
  \mathop{\max}\limits_{\tilde{\bm c}} & Save_{all} \left(\tilde{\bm c}\right) \\ 
  \text { s.t. }                       & \bm{1}^{T} \tilde{\bm c} \leq \tau  \enspace,\\ 
                                       & \tilde{\bm c} \geq \bm{0}
\end{array}$$
which will be solved by PGO. For completeness, we also consider the optimization problem with $Pay_{all}$ as the objective function, allowing a direct comparison between the results of the PGO method and those obtained from the Heuristic approach.

To illustrate our approach, we present a simple example using a simulated banking system. Specifically, we generate a system consisting of three banks and three illiquid assets, with one bank randomly selected to undergo bankruptcy. For other parameters, we adopt the settings from \citet{ma2021joint}. The total asset $w_i^0$ of Bank $i$ is randomly drawn from a uniform distribution $U(0.8, 1)$, and the liability network is modeled as an Erd\H{o}s-Rényi (E-R) network. The elasticity of illiquid assets is set to 5, while the remaining parameters describing the banking system are maintained as in the previous setup.
The rescue operation involves bailing out the two non-bankrupted banks. To fully present the rescue results, we vary the amount of rescue funds within the interval $[0, 1]$, randomly allocating the bailout amounts between the two surviving banks. This allows us to determine the values of $Pay_{all}$ and $Save_{all}$ in different bailout scenarios. Subsequently, we fit these results using neural networks with two hidden layers and five hidden layers, respectively.

\begin{figure}[ht]
\centering
\includegraphics[width=0.975\textwidth]{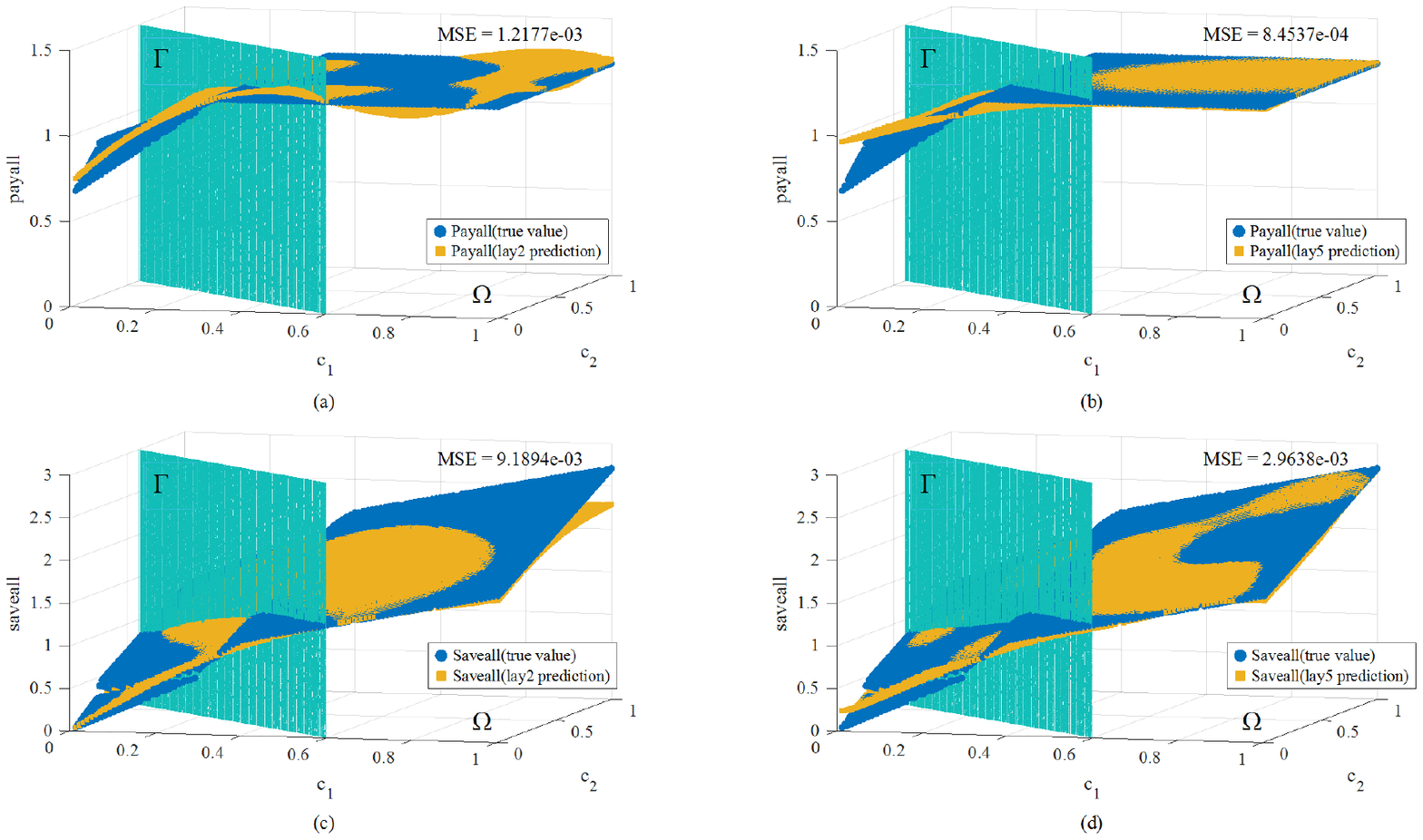}
    \caption{\small{\textbf{Example of the banking system with three banks.} 
    The diagram illustrates a three-bank financial system, where the first bank was rescued using $c_1$ funds, and the second bank was saved using $c_2$ funds. The third bank is omitted since it is bankrupt. $\Omega$ represents a plane, and $\Gamma$ is one of the vertical planes of $\Omega$. Each subplot displays two surfaces that depict the $Pay_{all}$ and $Save_{all}$ values of the banking system following the bailout approach, along with their corresponding fitted values. The vertical plane shows the scenario where the total fund is equivalent to the sum of funds used in the Heuristic method. In Panels (a) and (b), both of the blue surfaces indicate the true value of $Pay_{all}$ after bailout, while the yellow surfaces indicate the predicted values of $Pay_{all}$ obtained from the neural network with two and five hidden layers, respectively. In Panels (c) and (d), both of the blue surfaces indicate the true value of $Save_{all}$ after bailout, while the yellow surfaces indicate predicted values of $Save_{all}$ obtained from the neural network with two and five hidden layers, respectively.
    The MSE value displayed in the upper-left corner of each subplot represents the mean squared error between the true and estimated values of $Pay_{all}$ (or $Save_{all}$).
    }}
\label{fig:3banks}
\end{figure}

As shown in Figure~\ref{fig:3banks}, the true values of both $Save_{all}$ and $Pay_{all}$ illustrate that $Save_{all}$ is significantly more non-linear and harder to fit than $Pay_{all}$. This highlights the need for a deeper neural network to accurately model $Save_{all}$. 
The MSE values reported in Figure~\ref{fig:3banks} also reflect this pattern. A comparison between panels (a) and (b), or Panels~(c) and (d), shows that deeper neural networks lead to smaller MSEs when modeling $Pay_{all}$ or $Save_{all}$. Furthermore, comparing Panels~(a) with (c), or Panels~(b) with (d), reveals that $Save_{all}$ is inherently more difficult to approximate than $Pay_{all}$.
Furthermore, while $Pay_{all}$ reaches a plateau beyond a certain level of bailout funding, due to fixed total internal liabilities in the banking system, $Save_{all}$ continues to increase with the bailout fund. This discrepancy suggests that $Save_{all}$ serves as a more effective bailout target compared to $Pay_{all}$.
We further depict a $\Gamma$ plane, which is perpendicular to the $\Omega$ plane, representing the bailout budget from the Heuristic method. The intersection line between the true/predicted value plane and the $\Gamma$ plane illustrates the potential bailout effect corresponding to the specific bailout fund.

\subsubsection{Predetermined Budget} 
After determining our optimization objectives, we proceed to simulate the banking system. The number of banks, $n$, and the number of illiquid assets, $m$, are set to 10, 100, and 1000, respectively. To induce a shock in the system, we randomly select 10\% of the banks and bankrupt them. All other parameters remain unchanged from the previous setup. We then generate a training dataset by randomly creating rescue strategies and computing their corresponding $Pay_{all}$ and $Save_{all}$ values. To mitigate overfitting, we include some invalid samples that do not reflect real bailout scenarios, specifically cases where some banks in need of a bailout receive zero funds. The data set consists of 10,000 valid samples and 25,000 invalid ones. By including these samples, we ensure that the neural network does not erroneously learn that providing no bailout funds can lead to a greater bailout effect.

In this case, the bailout budget is predetermined. According to the heuristic principle, the execution of the heuristic algorithm generates the bailout amount, referred to as $budget$. Currently, we calculate the total loss of assets, $\tau_{max}$, incurred by the banking system during the shock. We then compare the bailout budget, $budget$, with the maximum asset loss, $\tau_{max}$, and set the lower value as the total bailout budget, denoted as $Budget$, i.e., $Budget = \min \left\{ budget, \tau_{max} \right\}$. The rationale behind this rule is that a bailout is impractical if the bailout amount, $budget$, exceeds the total assets lost, $\tau_{max}$.
It is worth noting that in the high-dimensional case ($n=1000$), the bailout budget, $budget$, generated by the Heuristic exceeds the total asset loss, $\tau_{max}$; thus, the Heuristic may be slightly sensitive to the dimensionality of the case.

Table~\ref{tb:one budget} presents the values $Pay_{all}$, $Save_{all}$, and $Ratio$ corresponding to the bailout vectors solved using different methods. The initial state represents the situation where the banking system has not yet received any bailout funds after the shock. The heuristic method prioritizes $Pay_{all}$ as its objective function. In contrast, when applying the PGO method, we use $Pay_{all}$ and $Save_{all}$ as objective functions, respectively. Based on these settings, we can obtain the corresponding indicators.

We begin by comparing the results of the heuristic method with those of the PGO, using $Pay_{all}$ as the objective function. In the low-dimensional case ($n=10$), both methods yield the same $Pay_{all}$ value of 5.600, but the corresponding $Save_{all}$ value for Heuristic is 0.644, which is lower than PGO's 0.719. In the medium-dimensional case ($n=100$), the performance of the PGO method is closer to that of Heuristic. In the high-dimensional case ($n=1000$), the PGO algorithm outperforms the Heuristic, with a rescue result of 303.902 compared to the Heuristic's 261.939. This indicates that the PGO algorithm is better suited for high-dimensional problems.
Furthermore, in both the $n=100$ and $n=1000$ cases, the $Save_{all}$ results of PGO, when $Save_{all}$ is used as the objective function, are higher than those obtained using $Pay_{all}$ as the objective function. This suggests that $Save_{all}$ is a more suitable indicator of rescue effectiveness than $Pay_{all}$. From an efficiency perspective, we observe that rescue operations are most efficient in the low-dimensional case ($n=10$), as reflected by the higher value $Ratio$ in this scenario compared to the other cases.

\begin{table}[H]\small
\centering
\renewcommand\arraystretch{1.25}
\caption{Results of the Heuristic and the PGO in \textit{Case~2} with predetermined budgets.}
\setlength{\tabcolsep}{5.35mm}{
\begin{tabular}{ccccc}
\toprule
\multicolumn{1}{c}{\textbf{Different Approach}} & \multicolumn{1}{l}{\textbf{Result}} & \multicolumn{1}{l}{$\bm{n=10}$} & \multicolumn{1}{l}{$\bm{n=100}$} & \multicolumn{1}{l}{$\bm{n=1000}$}    \\ 
\midrule
\multirow{2}{*}{Initial State}         
    & $Pay_{all}$   & 5.487  & 55.308     & 522.313             \\
    & $Budget$      & 0.076  & 3.080      & 54.237                   \\ 
\midrule
\multirow{3}{*}{\shortstack{Heuristic \\(Objective Function: $Pay_{all}$)}} 
    & $Pay_{all}$   & 5.600  & 59.765     & 590.471                  \\
    & $Save_{all}$  & 0.644  & 16.930     & 261.939                  \\
    & $Ratio$       & 8.472  & 5.497      & 4.830                    \\ 
\midrule
\multirow{3}{*}{\shortstack{PGO\\ (Objective Function: $Pay_{all}$)}}     
    & $Pay_{all}$   & 5.600  & 59.477     & 590.471                  \\
    & $Save_{all}$  
    & \textbf{0.719} 
    & 15.876   & \textbf{303.902}                  \\
    & $Ratio$       & \textbf{9.458}  & 5.155    & \textbf{5.603}                    \\ 
\midrule
\multirow{3}{*}{\shortstack{PGO\\(Objective Function: $Save_{all}$)}}    
    & $Pay_{all}$   & 5.600  & 59.420     & 590.471                  \\
    & $Save_{all}$  & \textbf{0.719}  & 16.525   & \textbf{307.152}    \\
    & $Ratio$       & \textbf{9.458}  & 5.365    & \textbf{5.663}       \\ 
\bottomrule
\end{tabular}}
\label{tb:one budget}
\end{table}

As demonstrated, the PGO algorithm proves to be effective in solving the optimization problem, even in the absence of an analytical expression for the objective function, thereby yielding superior results. In addition, it is recommended that the optimal bailout vector be determined based on the $Save_{all}$ objective, rather than the $Pay_{all}$ objective, which does not account for the price effects within the financial system.

\subsubsection{Flexible Budget} 
In the previous analysis, the amount of bailout funding was considered as $\min \left\{ budget, \tau_{max} \right\}$. However, in practice, it can be challenging for rescuers to determine the precise amount of funding required for a bailout. They may lack sufficient information to effectively balance the costs and benefits of the bailout. What rescuers may need more is an understanding of the different optimal relief outcomes corresponding to various levels of bailout funds, which would enable them to make more informed decisions. To address this challenge, we further develop a neural network model that represents the relationship between bailout effects and bailout funds. In this framework, the bailout fund range is set to $[0, \tau_{max}]$.

We can utilize the visual example presented in Figure~\ref{fig:3banks} to further illustrate the distinction. When the bailout constraint is known, the task is to fit the intersection line between the vertical plane, perpendicular to the $\Omega$ plane, and the surface where the intersection points satisfy the given bailout constraint. In contrast, when the bailout constraint is uncertain, the task is to fit the entire surface.

To evaluate the effectiveness of the PGO method, we assess its performance in ten different budget constraints. Table~\ref{tb:unknown budget} presents the $Save_{all}$ results derived from the PGO algorithm, as well as the optimal solutions obtained from the samples. This comparison is made between our method and the random sampling method. The results show that the solution produced by the PGO method aligns closely with the optimal sample solution, thus confirming the effectiveness of the approach, particularly in high-dimensional cases where $n=1000$.

\begin{table}[H]\small
\centering
\renewcommand\arraystretch{1.25}
\caption{Results of the Heuristic and the PGO in \textit{Case~2} with flexible budgets.}
\setlength\tabcolsep{2.25mm}{
\begin{tabular}{cccccccccc}
\toprule 
\multirow{2}{*}{} 
& \multicolumn{3}{c}{$\bm{n=10}$}   & \multicolumn{3}{c}{$\bm{n=100}$}   & \multicolumn{3}{c}{$\bm{n=1000}$}   \\ \cmidrule(r){2-4} \cmidrule(r){5-7} \cmidrule(r){8-10}
 & \multicolumn{1}{c}{\textbf{Sample}}& \multicolumn{1}{c}{\textbf{PGO}} & \multicolumn{1}{c}{\textbf{Ratio}}  & \multicolumn{1}{c}{\textbf{Sample}} & \multicolumn{1}{c}{\textbf{PGO}} & \multicolumn{1}{c}{\textbf{Ratio}}   & \multicolumn{1}{c}{\textbf{Sample}} & \multicolumn{1}{c}{\textbf{PGO}} & \multicolumn{1}{c}{\textbf{Ratio}} \\ \midrule
0.1$\tau_{max}$ 
& 0.851  &\textbf{0.875}   & 9.608
& 3.237  &\textbf{3.303}   & 3.619
& 31.801 &\textbf{32.095}   & 3.553 \\
0.2$\tau_{max}$ 
&1.278	&\textbf{1.278}	 & 7.021
&9.944	&\textbf{10.018}	 & 5.489
&63.603	&\textbf{63.931}    & 3.539 \\
0.3$\tau_{max}$ 
&1.369	&\textbf{1.369}	 & 5.014
&13.872	&\textbf{14.172}	 & 5.176
&100.962	&100.227                     & 3.698  \\
0.4$\tau_{max}$ 
&1.460	  &\textbf{1.460}	 & 4.010
&17.646	  &\textbf{19.079}	 & 5.227
&210.075  &\textbf{212.494} & 5.881  \\
0.5$\tau_{max}$
&1.551	  &1.551	      & 3.408
&21.202	  &20.895	      & 4.579
&257.300  &\textbf{257.760} & 5.707  \\
0.6$\tau_{max}$ 
&1.642	  &1.642       & 3.007
&23.700	  &23.354	   & 4.265
&280.043  &\textbf{281.901} & 5.201  \\
0.7$\tau_{max}$ 
&1.733	  &\textbf{1.733} & 2.720
&26.024	  &25.808	      & 4.040
&299.973  &\textbf{300.155} & 4.747  \\
0.8$\tau_{max}$ 
&1.824	  &\textbf{1.824}	 & 2.505
&28.107	  &27.931	& 3.826
&313.486  &\textbf{313.895} & 4.344 \\
0.9$\tau_{max}$ 
&1.915    &\textbf{1.915}	 & 2.338
&30.054	  &29.535	       & 3.596
&326.595  &\textbf{327.472} & 4.028 \\
$\tau_{max}$ 
&2.006	  &\textbf{2.006}  & 2.204
&31.145	  &30.940	           & 3.390
&340.400  &\textbf{340.812} & 3.773 \\
\bottomrule
\end{tabular}}
\label{tb:unknown budget}
\end{table}

\subsubsection{More Budget Constraints}
To further illustrate the advantages of the PGO method, consider the following example. Suppose that the total bailout funding is $0.1 \tau_{max}$, and the maximum bailout funds for each bank are given by $\displaystyle{\frac{\xi}{n_s} * 0.1 \tau_{max}}$, where $\xi$ is a parameter and $n_s$ represents the number of banks receiving bailout funds. In this example, we set $\xi = 1.5$. Since the bailout constraint has changed, it is necessary to regenerate the random sample and identify the optimal solution by ranking when using the sampling method. However, the flexibility of the PGO method allows us to update the constraints only during the optimization process. The results are presented in Table~\ref{tb:compare}.

\begin{table}[H]\small
\centering
\renewcommand\arraystretch{1.25}
\caption{Results of the random generation and the PGO in \textit{Case~2} with $0.1\tau_{max}$ budget and more constraints.}
\setlength\tabcolsep{5.5mm}{
\begin{tabular}{ccccccc}
\toprule 
& \multicolumn{2}{c}{$\bm{n=10}$} & \multicolumn{2}{c}{$\bm{n=100}$} & \multicolumn{2}{c}{$\bm{n=1000}$} \\ 
    \cmidrule(r){2-3} \cmidrule(r){4-5}  \cmidrule(r){6-7} 
& \textbf{Sample}  & \textbf{PGO} & \textbf{Sample} & \textbf{PGO}   & \textbf{Sample} & \textbf{PGO}  \\ 
\midrule
\multicolumn{1}{c}{$Save_{all}$}& 0.260  & \textbf{0.327} &5.037 & \textbf{6.209} & 31.793 & \textbf{32.094}          \\
\multicolumn{1}{c}{Time(s)} & 0.66 &0.87 &3.18 & 3.27 & 457.97         & \textbf{283.22} \\ 
\bottomrule
\end{tabular}}
\label{tb:compare}
\end{table}

Despite the relatively simple addition of constraints, in the high-dimensional case ($n=1000$), it takes 457.97 seconds to generate 100 samples that meet the basic constraints of the bailout fund. 
This sample size is relatively small for the case where $n=1000$. In other words, to achieve performance closer to the global optimum, the generation cost becomes substantial, and this cost must be incurred every time the constraints are changed. Thus, the method of generating samples and exhaustively searching for the optimal solution lacks flexibility.
In practical bailout scenarios, the time required increases as the constraints become more complex. Each time the constraints change, the samples must either be regenerated or re-screened from the existing data. However, the PGO method is not significantly affected by the complexity of the constraints. As long as the corresponding neural network is well trained, the constraints associated with the bailout can be flexibly adjusted, making the PGO method highly adaptable. In this case, the average solution time for PGO is 283.22 seconds. In scenarios where $n=10$ and $n=100$, the efficiency and performance of the PGO method are comparable to those of the randomly generated sample method, which further demonstrates the effectiveness of the PGO approach. As observed, the PGO method enhances flexibility in the solution process, particularly in high-dimensional cases.

\subsubsection{Optimal Bailout Ratio}\label{ap:ratio}
Upon observing that $Ratio$ changes synchronously with $Save_{all}$ in Section~\ref{sc:case2}, we conduct a further investigation into the properties of $Ratio$ and discover an interesting characteristic: As the budget changes, eventually a global maximum for $Ratio$ can be identified. Although rigorous proof is not provided, we observe the presence of optimal bailout efficiency across all three distinct dimensions of the banking system.

\begin{figure}[htbp]
    \centering
    \includegraphics[width=0.65\textwidth]{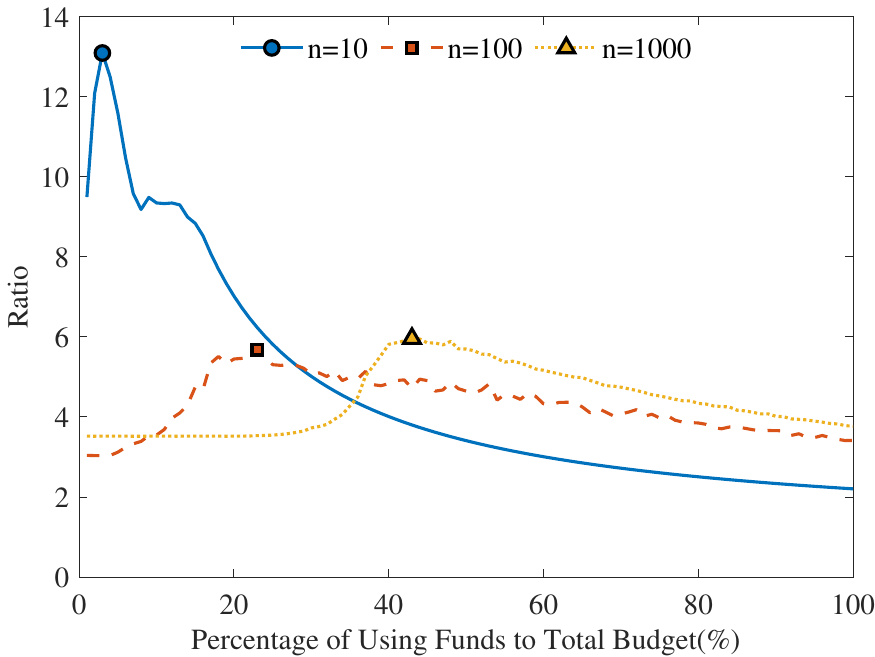}
    \caption{\small{\textbf{Optimal bailout ratios.} 
    This figure depicts the highest bailout ratio in a randomly generated sample with varying bailout funding constraints. The solid blue line represents the case of $n=10$, the dashed orange line represents the case of $n=100$, and the dotted yellow line represents the case of $n=1000$. The horizontal axis indicates the percentage of bailout funds used in relation to the total $\tau$, while the vertical axis represents the previously defined $Ratio$. The optimal bailout ratios for the $n=10$, $n=100$, and $n=1000$ cases are labeled with a circle, a square, and a triangle, respectively.
    }}
\label{fig:best_ratio}
\end{figure}

Following the data generation process described in Section~\ref{sc:case1&2} and the destruction of 10\% banks, we simulate various random rescue strategies with different bailout budgets and calculate the highest ratio for each budget. This allows us to obtain the optimal bailout efficiency within the sample. Figure~\ref{fig:best_ratio} presents the highest bailout ratio in a randomly generated sample under different bailout funding constraints in the extended E-N model. It is visually apparent that the efficiency curves for all three cases exhibit an inverted U-shape overall. Thus, we can adjust the bailout budget to achieve the highest rescue efficiency.

Although the shocks applied to all three banking systems are similar, we find that each system requires a different proportion of total funding to achieve its respective optimal bailout efficiency. As the size of the banking system increases, the required percentage of total funding to achieve optimal bailout efficiency also increases. This phenomenon, which mirrors the pattern of $Ratio$ in Table~\ref{tb:one budget}, indirectly illustrates how the complexity of the banking network amplifies the contagion of risk.

\section{Conclusions and Remarks}\label{sc:conclusion}
In this paper, we introduce a black-box optimization framework using the PGO method and apply it to solve the optimal bank bailout problem, which is crucial for mitigating the adverse effects of systemic risk. We establish a relationship between the rescue vector and its corresponding effect using neural networks, which are constructed on the basis of the fixed-point system within the extended E-N model. This approach enables the computation of the gradient of the objective function, thereby facilitating the optimization process.

Numerical comparisons demonstrate the effectiveness of PGO. In practical applications, PGO excels both in accuracy and computational efficiency, particularly in high-dimensional cases. In addition, PGO can be extended to address other systemic risk management challenges, such as optimizing the allocation of reserve funds. Furthermore, the PGO framework is not only suitable for solving the optimal bailout problem but also applicable to a wide range of non-traditional optimization problems.


\paragraph{Author Contributions}
S. H. Xiao has contributed to the methodology, conceptualization, computation, and original draft. J. L. Ma has contributed to the methodology, conceptualization, and writing. 
L. Xia has contributed to conceptualization, methodology, and writing.
S. S. Zhu has contributed to guidance, conceptualization, methodology, writing, and project administration.

\paragraph{Conflict of interest}
The authors declare that they have no conflict of interest.

\bibliographystyle{elsarticle-harv}
\bibliography{reference}

\newpage
\appendix
\setcounter{table}{0}
\setcounter{figure}{0}
\setcounter{equation}{0}
\setcounter{proposition}{0}

\renewcommand{\thetable}{C\arabic{table}}
\renewcommand{\thefigure}{A\arabic{figure}}
\renewcommand{\theequation}{A\arabic{equation}}
\renewcommand{\theproposition}{A\arabic{proposition}}
\renewcommand{\thesection}{A\arabic{section}}
\renewcommand{\thesubsection}{C\arabic{subsection}}
\renewcommand{\thesubsubsection}{A\arabic{subsubsection}}

\begin{appendices}
\newpage
\section{Derivatives of a Function Fitted by Neural Network}\label{ap:gradient} 
We begin by providing some preliminaries on the neural network.
Let $n_l$ denote the number of neurons in the $l^{th}$ layer and $p_{l}(\cdot)$ denote the activation function of the $l^{th}$ layer. The weight matrix of the $l^{th}$ layer $W^{\left[l\right]} \in \mathbb{R} ^{n_l \times n_{l-1}}$ comprises components $w_{ij}^{\left[l\right]}$, which correspond to the weight between the $j^{th}$ neurons of the $(l-1)^{th}$ layer and the $i^{th}$ neurons of the $l^th$ layer. 
Additionally, $\boldsymbol{b}^{\left[l\right]}=\left(b_{1}^{\left[l\right]}, b_{2}^{\left[l\right]}, \cdots, b_{n_{l}}^{\left[l\right]}\right)^{\top} \in \mathbb{R}^{n_{l}}$ represent the bias of the $l^{th}$ layer. Similarly, define $\boldsymbol{z}^{\left[l\right]}=\left(z_{1}^{\left[l\right]}, z_{2}^{\left[l\right]}, \cdots, z_{n_{l}}^{\left[l\right]}\right)^{\top} \in \mathbb{R}^{n_{l}}$ and $\boldsymbol{a}^{\left[l\right]}=\left(a_{1}^{\left[l\right]}, a_{2}^{\left[l\right]}, \cdots, a_{n_{l}}^{\left[l\right]}\right)^{\top} \in \mathbb{R}^{n_{l}}$ to represent the state vector and the output vector, respectively, of neurons in the $l^{th}$ layer.

The output of the neural network is $\boldsymbol{a}^{[L]}$, and according to the forward propagation, the process of the forward propagation is as follows:
$$\boldsymbol{x}
    =\boldsymbol{a}^{\left[1\right]} \rightarrow \boldsymbol{z}^{\left[2\right]} \rightarrow \cdots \rightarrow 
        \boldsymbol{a}^{\left[L-1\right]} \rightarrow \boldsymbol{z}^{\left[L\right]} \rightarrow \boldsymbol{a}^{\left[L\right]}
    =\boldsymbol{y},$$
where the input $\boldsymbol{x}$ is $n_0$ dimensional and the output $\boldsymbol{y}$ is $n_L$ dimensional.

Based on the back-propagation principle, we explore the process of obtaining the derivatives of a function fitted by a neural network. Denote the relationship between $\boldsymbol{x}$ and $\boldsymbol{y}$ as $\boldsymbol{y}=f\left(\boldsymbol{x}\right)$. Hence, the Jacobian matrix $J_{f(\boldsymbol{x})}$ that represents the derivative of $\boldsymbol{y}$ with respect to $\boldsymbol{x}$ is:
\begin{equation*}
    J_{f(\boldsymbol{x})} \triangleq\left(\begin{array}{ccc}
    \frac{\partial y_{1}}{\partial x_{1}}     & \cdots & \frac{\partial y_{1}}{\partial x_{n_{0}}} \\
    \vdots & \frac{\partial y_{u}}{\partial x_{v}}     & \vdots \\
    \frac{\partial y_{n_{L}}}{\partial x_{1}} & \cdots & \frac{\partial y_{n_L}}{\partial x_{n_{0}}}
    \end{array}\right)_{n_{L}\times n_{0}}.
\end{equation*}

According to the aforementioned mark conventions, the input $\boldsymbol{x}=\left(x_{1}, x_{2}, \cdots, x_{n_{0}}\right)^{\top}$ is equivalent to $z^{\left[0\right]}=\left(z_{1}^{\left[0\right]}, z_{2}^{\left[0\right]}, \dots,z_{n_0}^{\left[0\right]}\right)$. We therefore have $p_{0}^{\prime}\left(z_i^{\left[0\right]}\right)=1$.

Denote $\displaystyle{\frac{\partial y_{u}}{\partial z_{v}^{\left[l\right]}}}$ as $\displaystyle{\delta _{uv}^{\left[l\right]}}$, where $u=1, \cdots, n_{L}, v=1, \cdots n_{l}$. We have the following:
\begin{equation*}
    \delta_{u v}^{[l]}=\frac{\partial y_{u}}{\partial z_{v}^{[l]}}
    =\left\{\begin{array}{cl}
            \displaystyle{\sum_{n=1}^{n_{l+1}} \frac{\partial y_{u}}{\partial z_{n}^{[l+1]}} 
        \frac{\partial z_{n}^{[l+1]}}{\partial z_{v}^{[l]}}},   &\quad 0 \leq l \leq L-1 \\
            \left\{\begin{array}{cc}
                p_{L}^{\prime}\left(z_{v}^{[L]}\right) & u = v \\
                0 & u \neq v 
                \end{array}\right.
            & \quad l=L
    \end{array}\right. .
\end{equation*}

According to the following formula
$\displaystyle{\displaystyle{\frac{\partial z_n^{\left[l\right]}}{\partial z_v^{\left[l-1\right]}}
    =\frac{\partial z_n^{\left[l\right]}}{\partial a_v^{\left[l-1\right]}}\frac{\partial a_v^{\left[l-1\right]}}{\partial z_v^{\left[l-1\right]}}
    =w_{nv}^{\left[l\right]}p^{\prime}_{l-1}\left(z_v^{\left[l-1\right]}\right)}}$,
when $0 \leq l \leq L-1$, we can rewrite $\displaystyle{\delta_{uv}^{\left[l\right]}}$ as:
    $$\displaystyle{
    \delta_{uv}^{\left[l\right]}=\frac{\partial y_{u}}{\partial z_{v}^{\left[l\right]}}
                                =\sum_{n=1}^{n_{l+1}}\frac{\partial y_u}{\partial z_{n}^{\left[l+1\right]}} 
    w_{nv}^{\left[l+1\right]} p^{\prime}_{l}{\left(z_v^{\left[l\right]}\right)}, \quad 0\leq l\leq L-1.
    }$$
    
In particular, when $l=0$, there is:
\begin{equation}\label{eq:deltauv}
    \frac{\partial y_{u}}{\partial x_{v}} =\delta_{uv}^{\left[0\right]}
    =\frac{\partial y_{u}}{\partial z_{v}^{\left[ 0 \right]}}
    =\sum_{n=1}^{n_1}\frac{\partial y_u}{\partial z_n^{\left[1\right]}}\frac{\partial z_n^{\left[1\right]}}{\partial z_v^{\left[0\right]}}
    =\sum_{n=1}^{n_1}\delta _{un}^{\left[1\right]}w_{nv}^{\left[1\right]}p^{\prime}_0\left(z_{v}^{\left[0\right]}\right).
\end{equation}
Hence, according to Eq.(\ref{eq:deltauv}), we can calculate the gradient of $\boldsymbol{y}$ with respect to $\boldsymbol{x}$. 

We derived the above expression using the back-propagation principle. Applying the chain rule, we can obtain an essentially homogeneous formula through forward propagation as well.

Take $\displaystyle{\frac{\partial y_1}{\partial x_2}}$ as an example: 

\begin{equation*}
\begin{aligned}
    \frac{\partial y_{1}}{\partial x_{2}} 
    &=\frac{\partial y_{1}}{\partial z_{1}^{\left[L\right]}} \frac{\partial z_{1}^{\left[L\right]}}{\partial x_{2}} 
    \\ &
    =\frac{\partial y_{1}}{\partial z_{1}^{\left[L\right]}} \sum_{i=1}^{n_{L-1}} \frac{\partial z_{1}^{[L]}}{\partial a_{i}^{\left[L\right]}} \frac{\partial a_{i}^{\left[L-1\right]}}{\partial x_{2}} 
    \\ &
    =\frac{\partial y_{1}}{\partial z_{1}^{\left[L\right]}} \sum_{i=1}^{n_{L-1}} 
    \frac{\partial z_{1}^{\left[L\right]}}{\partial a_{i}^{\left[L-1\right]}}
    \frac{\partial a_{i}^{\left[L-1\right]}}{\partial z_{i}^{\left[L-1\right]}} 
    \frac{\partial z_{i}^{\left[L-1\right]}}{\partial x_{2}}
    \\ &
    =\frac{\partial y_{1}}{\partial z_{1}^{\left[L\right]}} \sum_{i=1}^{n_{L-1}} 
    \frac{\partial z_{1}^{\left[L\right]}}{\partial a_{i}^{\left[L-1\right]}}
    \frac{\partial a_{i}^{\left[L-1\right]}}{\partial z_{i}^{\left[L-1\right]}} 
    \left(
    \sum_{j=1}^{n_{L-2}}
    \frac{\partial z_{i}^{\left[L-1\right]}}{\partial a_{j}^{\left[L-2\right]}} 
    \frac{\partial a_{j}^{\left[L-2\right]}}{\partial x_2} 
        \right)
    \\ &
    =\frac{\partial y_{1}}{\partial z_{1}^{\left[L\right]}} \sum_{i=1}^{n_{L-1}} 
    \frac{\partial z_{1}^{\left[L\right]}}{\partial a_{i}^{\left[L-1\right]}}
    \frac{\partial a_{i}^{\left[L-1\right]}}{\partial z_{i}^{\left[L-1\right]}} 
    \left(
    \sum_{j=1}^{n_{L-2}}
    \frac{\partial z_{i}^{\left[L-1\right]}}{\partial a_{j}^{\left[L-2\right]}} 
    \frac{\partial a_{j}^{\left[L-2\right]}}{\partial z_{j}^{\left[L-2\right]}} 
    \frac{\partial z_{j}^{\left[L-2\right]}}{\partial x_2}
        \right).
\end{aligned}
\end{equation*}

According to 
$\displaystyle{\frac{\partial z_{i}^{\left[l\right]}}{\partial x_{2}} = \sum_{j=1}^{n_{l-1}}\frac{\partial z_{i}^{\left[l\right]}}{\partial a_{j}^{\left[l-1\right]}} \frac{\partial a_{j}^{\left[l-1\right]}}{\partial x_{2}} }$, we further have:
\begin{footnotesize}\begin{equation}\label{eq:ytox1}
\begin{aligned}
    \frac{\partial y_{1}}{\partial x_{2}} =\frac{\partial y_{1}}{\partial z_{1}^{\left[L\right]}} \sum_{i=1}^{n_{L-1}} 
    \frac{\partial z_{1}^{\left[L\right]}}{\partial a_{i}^{\left[L-1\right]}}
    \frac{\partial a_{i}^{\left[L-1\right]}}{\partial z_{i}^{\left[L-1\right]}} 
    \left\{\sum_{j=1}^{n_{L-2}}
    \frac{\partial z_{i}^{\left[L-1\right]}}{\partial a_{j}^{\left[L-2\right]}} 
    \frac{\partial a_{j}^{\left[L-2\right]}}{\partial z_{j}^{\left[L-2\right]}} 
    \left[\sum_{k=1}^{n_{L-3}}
    \frac{\partial z_{j}^{\left[L-2\right]}}{\partial a_{k}^{\left[L-3\right]}} 
    \frac{\partial a_{k}^{\left[L-3\right]}}{\partial z_{k}^{\left[L-3\right]}} 
    \dots
    \left(\sum_{q=1}^{n_{1}}
    \frac{\partial z_{p}^{\left[2\right]}}{\partial a_{q}^{\left[1\right]}} 
    \frac{\partial a_{q}^{\left[1\right]}}{\partial z_{q}^{\left[1\right]}} 
    \frac{\partial a_{q}^{\left[1\right]}}{\partial x_{2}} 
    \right)\right]\right\}.
\end{aligned}\end{equation}\end{footnotesize}
Based on $\displaystyle{\frac{\partial a_{i}^{[l]}}{\partial z_{i}^{[l]}}=p_{l}^{\prime}\left(z_{i}^{[l]}\right)}$ and $\displaystyle{\frac{\partial z_{i}^{[l]}}{\partial a_{j}^{[l-1]}}=w_{i j}^{[l]}}$, we can rewrite Eq.(\ref{eq:ytox1}) as:
\begin{footnotesize}\begin{equation*}\begin{aligned}
    \frac{\partial y_{1}}{\partial x_{2}} =p_L^{\prime}\left(z_1^{\left[L\right]}\right)\sum_{i=1}^{n_{L-1}}w_{1i}^{\left[L\right]}p_{L-1}^{\prime}\left(z_i^{\left[L-1\right]}\right)
    \left\{
        \sum_{j=1}^{n_{L-2}}w_{ij}^{\left[L-1\right]}p_{L-2}^{\prime}\left(z_j^{\left[L-2\right]}\right)
        \left[
            \sum_{k=1}^{n_{L-3}}w_{jk}^{\left[L-2\right]}p_{L-3}^{\prime}\left(z_k^{\left[L-3\right]}\right)
            \dots
            \left(
                \sum_{q=1}^{n_{1}}w_{pq}^{\left[2\right]}p_{1}^{\prime}\left(z_q^{\left[1\right]}\right)w_{q2}^{\left[1\right]}
        \right)\right]\right\}.
\end{aligned}\end{equation*}\end{footnotesize}

Hence, it is easy to further obtain:
\begin{footnotesize}
\begin{equation*}\begin{aligned}
\frac{\partial y_{u}}{\partial x_{v}} &=p_L^{\prime}\left(z_u^{\left[L\right]}\right)\sum_{i=1}^{n_{L-1}}w_{ui}^{\left[L\right]}p_{L-1}^{\prime}\left(z_i^{\left[L-1\right]}\right)
\left\{
    \sum_{j=1}^{n_{L-2}}w_{ij}^{\left[L-1\right]}p_{L-2}^{\prime}\left(z_j^{\left[L-2\right]}\right)
    \left[
        \sum_{k=1}^{n_{L-3}}w_{jk}^{\left[L-2\right]}p_{L-3}^{\prime}\left(z_k^{\left[L-3\right]}\right)
        \dots
        \left(
            \sum_{q=1}^{n_{1}}w_{pq}^{\left[2\right]}p_{1}^{\prime}\left(z_q^{\left[1\right]}\right)w_{qv}^{\left[1\right]}
\right)\right] \right\},
\end{aligned}\end{equation*}
\end{footnotesize}
which can be simplified by the back-propagation principle and has been described in Eq.(\ref{eq:deltauv}).

\newpage
\section{Gradient Projection Algorithm}\label{ap:gradient projection}
We briefly introduce the application of GPA in this subsection. According to \cite{rosen1960gradient}, we consider the following optimization problem:
$$ \begin{array}{ll}
    \min & f(\boldsymbol{x}) \\ 
    \text {s.t.} & \boldsymbol{A} \boldsymbol{x} \geqslant \boldsymbol{b} \enspace, \\ 
                    & \boldsymbol{E} \boldsymbol{x}=\boldsymbol{e} 
\end{array}$$
where $f(\boldsymbol{x})$ is a differentiable function and both of $\boldsymbol{A}$ and $\boldsymbol{E}$ are $m\times n$ dimensional matrices.

The GPA is constructed as follows:

\vspace{0.3cm}
{\bf  Gradient Projection Algorithm} 
{\it\small \begin{description}
    \item [Step 1:] Set the initial feasible point $\boldsymbol{x}^{(k)}$, $k=:0$;
    \item [Step 2:] Calculate $\boldsymbol{A} \boldsymbol{x}^{(k)}$ and decompose $\boldsymbol{A}$ and $\boldsymbol{b}$ into $ \left[\begin{array}{c}\boldsymbol{A}_{1} \\ \boldsymbol{A}_{2}\end{array}\right] and \left[\begin{array}{l}\boldsymbol{b}_{1} \\ \boldsymbol{b}_{2}\end{array}\right] $ such that $ \boldsymbol{A}_{1} \boldsymbol{x}^{(k)}=\boldsymbol{b}_{1}, \boldsymbol{A}_{2} \boldsymbol{x}^{(k)}>\boldsymbol{b}_{2}$;
    \item [Step 3:] Let $ \boldsymbol{M}=\left[\begin{array}{l}\boldsymbol{A}_{1} \\ \boldsymbol{E}\end{array}\right] $;
    
    If $ \boldsymbol{M}$ is empty, let $ \boldsymbol{P}= \boldsymbol{I}$, where $\boldsymbol{I}$ is an identity matrix; else, let $ \boldsymbol{P}=\boldsymbol{I}-\boldsymbol{M}^{\mathrm{T}}\left(\boldsymbol{M} \boldsymbol{M}^{\mathrm{T}}\right)^{-1} \boldsymbol{M} $;
    \item [Step 4:] Let $ \boldsymbol{d}^{(k)}=-\boldsymbol{P} \boldsymbol{\nabla} f\left(\boldsymbol{x}^{(k)}\right) $. If $\boldsymbol{d}^{(k)} \neq \bm{0} $ , turn to Step 6; else, turn to Step 5;
    \item [Step 5:] If $\boldsymbol{M}$ is empty, stop; else, let $ \boldsymbol{W}=\left(\boldsymbol{M} \boldsymbol{M}^{T}\right)^{-1} \boldsymbol{M} \boldsymbol{\nabla} f\left(\boldsymbol{x}^{(k)}\right)=\left[\begin{array}{l}\boldsymbol{u} \\ \boldsymbol{v}\end{array}\right] $;
    
    If $\boldsymbol{u}\geq 0$, stop, $\boldsymbol{x}^{(k)}$ is the K-T point; else, select a negative component $u_j$ and correct the row corresponding to $u_j$ in $\boldsymbol{A}_{1}$, go back to Step 3.
    \item [Step 6:] Calculate $ \lambda_{\max}=\left\{\begin{array}{ll}\min \left\{\frac{\dot{b}_{i}}{\hat{d}_{i}} \mid \hat{d}_{i}<0\right\}, & \hat{\boldsymbol{d}} \geqslant \bm{0} \\ \infty, & \hat{\boldsymbol{d}} \geqslant \bm{0}\end{array} \right.$;
    
    Search gradually along 0 $\longrightarrow $ $\lambda_{\max}$ and select $\lambda$ in the interval $\left[0,\lambda_{\max}\right]$;
    
    Let $ \boldsymbol{x}^{(k+1)}=\boldsymbol{x}^{(k)}+\lambda_{k} \boldsymbol{d}^{(k)} $ and set $k=:k+1$, go back to Step 2.
\end{description}}

Notably, in \cite{rosen1960gradient}, we can attain the optimal step size at {\bf Step 6} of GPA by solving the following optimization problem:
$$ \begin{array}{ll}
    \min & f\left(\boldsymbol{x}^{(k)}+\lambda \boldsymbol{x}^{(k)}\right) \\ \text { s.t. } & 0 \leqslant \lambda \leqslant \lambda_{\max }. 
\end{array}$$
However, because the objective function of the bailout problem lacks an analytic form, we opt to select step sizes available within the interval $\left[0,\lambda_{\max}\right]$ instead of pursuing the optimal step size. Specifically, we divide the interval $\left[0,\lambda_{\max}\right]$ equally into $N$ intervals. Starting from the first subinterval to the last subinterval, we compute the function value of the endpoint of each subinterval. As long as the value of $f\left(\boldsymbol{x}^{(k)}+\lambda_{i} \boldsymbol{x}^{(k)}\right)-f\left(\boldsymbol{x}^{(k)}\right)$ is greater than $\epsilon$ which is a parameter we set, we choose $\lambda_{i}$ as the step size. This method is crude, but effective. 

\section{Supplemental Materials}
In this appendix, we begin by demonstrating the effectiveness of the PGO method. Afterwards, we provide the structural information of the neural networks used in this paper.
\subsection{Effectiveness of PGO Algorithm}\label{ap:slsqp}
In this subsection, we demonstrate the effectiveness of the PGO method. As demonstrated in Section \ref{sub:ENLP}, we can establish a connection between the LP and the clearing payment vector in the E-N model. If we opt to utilize the PGO method, the corresponding linear program is:
$$\begin{array}{ll}
\max \limits_{\boldsymbol{l}} & \bm{1}^{T} \boldsymbol{l} \\
        \text { s.t. } & \boldsymbol{c}+\Pi^{T} \boldsymbol{l} \geq \boldsymbol{l} \enspace. \\
                        & \bm{0} \leq \boldsymbol{l} \leq \overline{\boldsymbol{l}}
\end{array}$$

We first use LINPROG implemented in MATLAB to solve the above programming problem and obtain the standard solution, $\bm l^{*}$. Then we apply the PGO approach to solve the same programming problem and compare the results. Specifically, we randomly generate $\bm l_{i}$ samples and compute their corresponding objective function value, $\bm{1}^{T} \boldsymbol{l}$, to obtain the sample for training. 
Table \ref{tb:LPprogram} shows the results obtained, with content and structure consistent with those of Table \ref{tb:LP-rescue}.
Using only a small number of hidden layers in a neural network, we can achieve an accurate solution to this special LP problem. The result demonstrates the effectiveness of our approach.
\begin{table}[H]\small
\centering
\renewcommand\arraystretch{1.25}
\caption{Results of LINPROG and PGO in \textit{Case 1}.}
\setlength{\tabcolsep}{7.5mm}{
\begin{tabular}{ccccc}
\toprule
\multirow{2}{*}{$\bm{n}$} & \multirow{2}{*}{\textbf{LINPROG}}     & \multicolumn{3}{c}{\textbf{PGO-SLSQP}} \\ \cline{3-5} 
&                         & \bm{$lay=2$} & \bm{$lay=3$}    & \bm{$lay=4$}        \\ 
\midrule
\multirow{2}{*}{$n=10$}   & \multirow{2}{*}{40.515}   & 40.511    & \textbf{40.515}    & \textbf{40.515}        \\
                            &                           & 99.99\%   & 100.00\%  & 100.00\%  \\ 
\midrule
\multirow{2}{*}{$n=100$}  & \multirow{2}{*}{513.537}  & \textbf{513.537}    & \textbf{513.537}   & \textbf{513.537}      \\
                            &                          & 100.00\%   & 100.00\%   & 100.00\%  \\ 
\midrule
\multirow{2}{*}{$n=1000$} & \multirow{2}{*}{5572.189} & \textbf{5572.189}  & \textbf{5572.189}  & \textbf{5572.189}    \\
                            &                           & 100.00\%  & 100.00\%  & 100.00\%  \\ 
\bottomrule
\end{tabular}}
\label{tb:LPprogram}
\end{table}

\subsection{Construction Information of Neural Networks}\label{ap:inf-nn}
Table \ref{tb:network-detail} provides the hidden layer architectures of the neural networks used in \textit{Case 1} and \textit{Case 2} in this study. Specifically, the column ``Neural Neuron'' lists the number of neurons used in the hidden layers of neural networks, and the column ``Activation Function'' lists the activation functions used in the hidden layers, where ``Sigmoid'' refers to the commonly used sigmoid function, and ``ReLU'' refers to the commonly used Rectified Linear Unit function. Furthermore, ``OBJ FUNC'' is the abbreviation of ``objective function'', which is used for brevity.

\begin{sidewaystable}\small
    \renewcommand\arraystretch{1.25}
    \captionsetup{font={small}}
    \caption{Architectures of neural networks used in two practical cases.}
        \centering
    \begin{tabular}{cllllllllllll}
    \toprule
    \multicolumn{3}{c}{\textbf{Case}}    & \multicolumn{5}{c}{\textbf{Neural Neuron}} & \multicolumn{5}{c}{\textbf{Activation Function}} \\ \midrule
    \multirow{12}{*}{\textit{Case 1}}       
        & $n=10$  & $lay=2$ & 256   & 128  &      &      &      & Sigmoid    & Sigmoid    &        &       &      \\
        & $n=10$  & $lay=3$ & 128   & 64   & 32   &      &      & Sigmoid    & Sigmoid    & Sigmoid    &       &      \\
        & $n=10$  & $lay=4$ & 256   & 128  & 64   & 32   &      & ReLU   & ReLU   & ReLU   & Sigmoid   &      \\
        & $n=10$  & $lay=5$ & 128   & 64   & 32   & 16   & 4    & Sigmoid    & Sigmoid    & Sigmoid    & Sigmoid   & Sigmoid  \\
        & $n=100$ & $lay=2$ & 128   & 64   &      &      &      & ReLU   & Sigmoid    &        &       &      \\
        & $n=100$ & $lay=3$ & 128   & 64   & 32   &      &      & ReLU   & ReLU   & Sigmoid    &       &      \\
        & $n=100$ & $lay=4$ & 256   & 128  & 64   & 32   &      & Sigmoid    & Sigmoid    & Sigmoid    & Sigmoid   &      \\
        & $n=100$ & $lay=5$ & 128   & 64   & 32   & 16   & 4    & ReLU   & ReLU   & ReLU   & ReLU  & Sigmoid  \\
        &$n=1000$ & $lay=2$ & 256   & 128  &      &      &      & Sigmoid    & Sigmoid    &        &       &      \\
        &$n=1000$ & $lay=3$ & 128   & 64   & 32   &      &      & ReLU   & ReLU   & Sigmoid    &       &      \\
        &$n=1000$ & $lay=4$ & 256   & 128  & 64   & 32   &      & Sigmoid    & Sigmoid    & Sigmoid    & Sigmoid   &      \\
        &$n=1000$ & $lay=5$ & 128   & 64   & 32   & 16   & 4    & Sigmoid    & Sigmoid    & Sigmoid    & Sigmoid   & Sigmoid  \\ \hline
    \multirow{8}{*}{\begin{tabular}[c]{@{}c@{}}\textit{Case 2} \\ Predetermined Budget\end{tabular}}   
        & $n=10$  & OBJ FUNC: $Pay_{all}$  
        & 180   & 120 &   &  &  & Sigmoid & Sigmoid &     &  &      \\
        & $n=10$  & OBJ FUNC: $Save_{all}$ 
        & 180   & 120  &  & &   & Sigmoid & Sigmoid & & &      \\
        & \multirow{2}{*}{$n=100$} & \multirow{2}{*}{OBJ FUNC: $Pay_{all}$}  
        & 200   & 200  & 200 & 200  & 200  & Sigmoid  & Sigmoid  & Sigmoid  & Sigmoid & Sigmoid  \\
        & &     & 200  & 200 & 200  & 200  & 5    & Sigmoid  & Sigmoid  & Sigmoid & Sigmoid & Sigmoid  \\
        & \multirow{2}{*}{$n=100$} & \multirow{2}{*}{OBJ FUNC: $Save_{all}$} 
        &      200  & 200  & 200 & 200  & 200  & Sigmoid  & Sigmoid  & Sigmoid  & Sigmoid & Sigmoid  \\
        & &  & 200  & 200 & 200  & 200  & 5    & Sigmoid  & Sigmoid  & Sigmoid & Sigmoid & Sigmoid  \\
        & $n=1000$ & OBJ FUNC: $Pay_{all}$  
        & 480   & 240  & & & & Sigmoid  & Sigmoid  &  &   &      \\
        & $n=1000$ & OBJ FUNC: $Save_{all}$ 
        & 480   & 240  & & & & Sigmoid  & Sigmoid  &  &   &      \\ \hline
    \multirow{4}{*}{\begin{tabular}[c]{@{}c@{}}\textit{Case 2} \\ Flexible Budget\end{tabular}} 
        & $n=10$  & OBJ FUNC: $Save_{all}$  
        & 180   & 120  &     &  &   & Sigmoid  & Sigmoid  &      &     &    \\
        & \multirow{2}{*}{$n=100$} & \multirow{2}{*}{OBJ FUNC: $Save_{all}$} 
        & 200   & 200  & 200 & 200  & 200  & Sigmoid  & Sigmoid  & Sigmoid & Sigmoid & Sigmoid  \\
        &   &   & 200  & 200 & 200  & 200  & 5    & Sigmoid  & Sigmoid & Sigmoid & Sigmoid & Sigmoid  \\
        &$n=1000$ & OBJ FUNC: $Save_{all}$   
        & 1500  & 500  & & & & ReLU & Sigmoid  & & &      \\ 
        \bottomrule
\end{tabular}
\label{tb:network-detail}
\end{sidewaystable}

\end{appendices}

\end{document}